\documentclass[11pt]{article}
\usepackage{graphicx,array,epic,amscd,amssymb}
\usepackage[totalwidth=480pt,totalheight=680pt]{geometry}
\usepackage{amsmath}

\newcommand{\DIS}{\displaystyle}

\newcommand{\qed}{\hfill $\Box$}
\def\C{{\mathbb C}}
\def\Z{{\mathbb Z}}
\def\R{{\mathbb R}}
\def\N{{\mathbb N}}

\def\nn{{\nonumber}}
\def\bLambda{\text{\mathversion{bold}{$\Lambda$}}}

\newtheorem{theorem}{Theorem}
\newtheorem{corollary}{Corollary}

\newtheorem{example}{Example}
\newtheorem{lemma}{Lemma}
\newtheorem{proposition}{Proposition}
\newtheorem{definition}{Definition}

\begin{document}

\title{\bf A geometric realization of the periodic discrete Toda lattice and its tropicalization}

\author{Atsushi \sc{Nobe}\footnote{nobe@faculty.chiba-u.jp}\\
\normalsize{Department of Mathematics, Faculty of Education, Chiba University,}\\
\normalsize{1-33 Yayoi-cho Inage-ku, Chiba 263-8522, Japan}
}
\date{}
\maketitle

\begin{abstract}
An explicit formula concerning curve intersections equivalent to the time evolution of the periodic discrete Toda lattice is presented.
First, the time evolution is realized as a point addition on a hyperelliptic curve, which is the spectral curve of the periodic discrete Toda lattice, then the point addition is translated into curve intersections. 
Next, it is shown that the curves which appear in the curve intersections are explicitly given by using the conserved quantities of the periodic discrete Toda lattice.
Finally, the formulation is lifted to the framework of tropical geometry, and a tropical geometric realization of the periodic box-ball system is constructed via tropical curve intersections.
\end{abstract}


\section{Introduction}
In recent years, many studies on ultradiscrete integrable systems, or integrable cellular automata have intensively been made by using various mathematical tools such as combinatorics \cite{MSTTT97,FOY00,YYT03,NY04}, ultradiscretization procedure \cite{TTMS96,NT98,TTM00,IMNS04}, crystal bases of quantum groups \cite{HIK99,HHIKTT01,HKOTY02,IKT12}, and so forth.
In particular, among many studies on geometric aspects of ultradiscrete integrable systems, several remarkable works have successively been made by Kimijima-Tokihiro \cite{KT02}, Inoue-Takenawa \cite{IT08,IT09}, and Iwao \cite{Iwao10} in terms of tropical geometry and ultradiscretization procedure.
They have solved the initial value problems to box-ball systems with periodic boundary conditions (abbreviated as periodic box-ball systems or pBBS shortly) by using tropical hyperelliptic curves, their tropical Jacobians, ultradiscrete theta functions, and ultradiscretization of Abelian integrals.
Through their works, we have gradually recognized the advantage in applying the method of tropical geometry to ultradiscrete integrable systems.
Therefore, when we focused our interest on the geometric aspect of ultradiscrete integrable systems we usually chose the framework of tropical geometry as the analysis tool \cite{Nobe08, KNT08,KKNT09,Nobe11,IKT12}. 

In this paper, we will go a step further and try to realize a pBBS in the framework of tropical geometry more precisely.
Here a `precise' geometric realization means that we will give the time evolution of the pBBS not by linear flows on the tropical Jacobian but by tropical curve intersections.
If we try to do so we immediately notice that the task is not so easy, because to find tropical curve intersections equivalent to the time evolution of the pBBS is essentially the same as to obtain the general solution to a simultaneous system of piecewise linear equations. 
(Note that we do not have a Cramer-type formula for such a system of equations.)
However, preceding studies on ultradiscrete systems shed light on us and we found a short cut.
We first establish a geometric realization of the periodic discrete Toda lattice (abbreviated as pdTL), whose ultradiscretization gives the pBBS, by using curve intersections.
We then apply the ultradiscretization procedure to the members in the curve intersections, and finally obtain a geometric realization of the pBBS in terms of tropical curve intersections.

To carry out our plan, this paper is organized as follows. 
We first review the basic results concerning addition of points on hyperelliptic curves in section \ref{sec:AOH}.
In section \ref{sec:GRPDTL}, we then introduce the pdTL and investigate its spectral curve and the conserved quantities.
In section \ref{sec:grpdtl}, we establish curve intersections equivalent to the time evolution of the pdTL.
We then focus on ultradiscrete systems.
In section \ref{sec:ATHCPBS}, we introduce the ultradiscrete periodic Toda lattice (UD-pTL) and its spectral curve, and review addition of points on tropical hyperelliptic curves.
(If we take the initial values of the UD-pTL in positive integers then we obtain the pBBS.)
Finally, in section \ref{sec:AGROP}, we apply the procedure of ultradiscretization to the rational functions which appear in the geometric realization of the pdTL.
Then the geometric framework of the pdTL can naturally be translated into that of tropical geometry, and we obtain a geometric realization of the UD-pTL by using tropical curve intersections.

\section{Addition on hyperelliptic curves}
\label{sec:AOH}
We first present a brief review of addition of points on hyperelliptic curves (see for example \cite{MWZ98}).
\subsection{Hyperelliptic curves}
Let $h(x)$ be the monic polynomial of degree $2g+2\geq4$:
\begin{align*}
h(x)
=
x^{2g+2}
+
a_{2g+1}x^{2g+1}
+
a_{2g}x^{2g}
+
\cdots
+
a_1x
+
a_0,
\end{align*}
where $a_0,a_1,\ldots,a_{2g+1}\in\C$.
Consider the affine curves $H_0$ on $\C_{(x,y)}^2$ and $H_\infty$ on $\C_{(u,v)}^2$:
\begin{align*}
&H_0
:=
\left(
y^2-h(x)=0
\right),\quad
&H_\infty
:=
\left(
v^2-u^{2g+2}h\left(\frac{1}{u}\right)=0
\right),
\end{align*}
where $(x,y)$ and $(u,v)$ are the standard coordinates of $\C_{(x,y)}^2$ and $\C_{(u,v)}^2$, respectively.
Assume that $h(x)=0$ has no multiple root and hence the both affine curves are non-singular.
Let the projections on the curves be
\begin{align*}
\pi_0:H_0\to\C_x;\ (a,b)\mapsto a,\quad
\pi_\infty:H_\infty\to\C_u;\ (\alpha,\beta)\mapsto \alpha.
\end{align*}
By gluing $H_0$ with $H_\infty$ in terms of the bi-holomorphic map $H_0\setminus\pi_0^{-1}(0)\to H_\infty\setminus\pi_\infty^{-1}(0)$;
\begin{align*}
(x,y)
\mapsto
(u,v)=\left(
\frac{1}{x},
\frac{y}{x^{g+1}}
\right),
\end{align*}
we obtain the hyperelliptic curve $H:=H_0\cup H_\infty$ of genus $g$.

Substitute $x=1/t$ into $y^2-h(x)=0$ then we obtain
\begin{align*}
\left(t^{g+1}y\right)^2
=
1+a_{2g+1}t+a_{2g}t^2+\cdots+a_1t^{2g+1}+a_0t^{2g+2}.
\end{align*}
If $t$ goes to 0 this equation reduces to $\left(t^{g+1}y\right)^2\sim1$.
Hence we obtain $(x,y)\sim\left({1}/{t},\pm{1}/{t^{g+1}}\right)$, or equivalently $(u,v)\sim\left(t,\pm1\right)$ for sufficiently small $t$.
Therefore, the curve $H$ can be expressed as
\begin{align*}
H
=
H_0
\cup
\left\{
P_\infty,P_\infty^\prime
\right\},
\end{align*}
where $P_\infty=(0,1), P_\infty^\prime=(0,-1)\in\C_{(u,v)}^2$ are the points at infinity.

Consider the involution $\iota:H\to H; P=(x,y)\mapsto P^\prime=(x,-y)$.
We call $P^\prime$ the conjugate of $P$.

Let $\mathcal{D}(H)$ be the divisor group of $H$.
If $D\in\mathcal{D}(H)$ then $D$ has the form $D=\sum_{P\in H}a_PP$, where $a_P\in\Z$ and $a_P=0$ except for finite $P$.
The number $\sum_{P\in H}a_P$ is called the degree of $D$ and is denoted by $\deg D$.
If $a_P\geq 0$ for all $P\in H$ the divisor $D$ is called effective and is denoted by $D>0$.

\subsection{Canonical maps}
Let $\mathcal{D}_0(H):=\left\{D\in\mathcal{D}(H)\ |\ \deg D=0\right\}$ be the group of divisors of degree 0 on $H$.
Also let $\mathcal{D}_l(H):=\left\{(k)\ |\ k\in \C(H)\right\}$ be the group of principal divisors of rational functions on $H$, where $\C(H)$ is the field of rational functions on $H$ and $(k)$ denotes the principal divisor of $k\in\C(H)$.
We define the Picard group ${\rm Pic}^0(H)$ to be the residue class group ${\rm Pic}^0(H):=\mathcal{D}_0(H)\slash\mathcal{D}_l(H)$.

Let $\mathcal{D}_g^+(H):=\left\{D\in\mathcal{D}(H)\ |\ D>0,\deg D=g\right\}$ be the set of effective divisors of degree $g$ on $H$.
For simplicity, we denote the element $P_{1}+P_{2}+\cdots+P_{g}$ of $\mathcal{D}_{g}^{+}(H)$ by $D_P$. 

Fix an element $D^\ast$ of $\mathcal{D}_g^+(H)$.
Define the canonical map $\Phi:\mathcal{D}_g^+(H)\to {\rm Pic}^0(H)$ to be
\begin{align*}
\Phi(A)
:\equiv
A-D^\ast
\quad
\mbox{(mod $\mathcal{D}_l(H)$)}
\qquad
\mbox{for $A\in\mathcal{D}_g^+(H)$.}
\end{align*}
The following theorem is easily shown.
\begin{theorem}\label{thm:phisurjec}
The canonical map $\Phi$ is surjective.
In particular, $\Phi$ is bijective if $g=1$.
\end{theorem}

(Proof)\quad
Let $D$ be an arbitrary element of $\mathcal{D}_0(H)$.
By the Riemann-Roch theorem, we have
\begin{align*}
\dim L\left(D+D^\ast\right)
=
\deg\left(D+D^\ast\right)+1-g+\dim L\left(W-D-D^\ast\right)
\geq1,
\end{align*}
where $L(D):=\{k\in \C(H)\ |\ (k)+D>0\}$ and $W$ is a canonical divisor.
This inequality implies that there exists a rational function $h$ on $H$ such that $(h)+D+D^\ast>0$.
If we put $A=(h)+D+D^\ast$ then $A$ is an effective divisor of degree $g$, and hence $A\in\mathcal{D}_g^+(H)$.
Moreover, we have
\begin{align*}
\Phi(A)
\equiv D
\quad
\mbox{(mod $\mathcal{D}_l(H)$)}.
\end{align*}
Therefore, the map $\Phi$ is surjective.

Now assume $g=1$.
If $\Phi(P)=\Phi(Q)$ holds then $P\equiv Q$ (mod $\mathcal{D}_l(H)$), and hence there exists a rational function $h$ on $H$ such that $(h)=P-Q$.
Since $P=(h)+Q$ is effective, we find $h\in L(Q)$.
Note that the degree of $Q$ satisfies $\deg Q=1>2g-2$.
By the Riemann-Roch theorem, we have
\begin{align*}
\dim L(Q)=1.
\end{align*}
This means that $L(Q)$ is spanned by a constant function, say $1$; $L(Q)=\langle 1\rangle$.
Therefore, $P-Q=(1)=0$ holds.
Thus we conclude that the map $\Phi$ is injective.
\qed

\subsection{Addition on symmetric products}
By using the surjection $\Phi$, we induce addition of points on the $g$-th symmetric product ${\rm Sym}^g(H):=H^g/\mathfrak{S}_g$ from ${\rm Pic}^0(H)$.

There exists a bijection
\begin{align*}
\mu:\mathcal{D}_{g}^{+}(H)\to{\rm Sym}^g(H);\ 
D_P\mapsto
\mu(D_P)
=
\left\{P_1,P_2,\cdots,P_g\right\}.
\end{align*}
We denote $\mu(D_P)\in{\rm Sym}^g(H)$ by $d_P$.
Put $\tilde\Phi:=\Phi\circ\mu^{-1}:{\rm Sym}^g(H)\to {\rm Pic}^0(H)$.
Note that we have ${\rm Pic}^0(H)=\left\{\tilde\Phi(d)\ |\ d\in {\rm Sym}^g(H)\right\}$ because $\tilde\Phi$ is surjective.
For $d_P, d_Q\in{\rm Sym}^g(H)$, we define $d_P\oplus d_Q$ to be an element in the subset
\begin{eqnarray*}
\tilde\Phi^{-1}\left(\tilde\Phi(d_P)+\tilde\Phi(d_Q)\right)
\subset{\rm Sym}^g(H).
\end{eqnarray*}
Since $\tilde\Phi$ is surjective, 
$\tilde\Phi^{-1}\left(\tilde\Phi(d_P)+\tilde\Phi(d_Q)\right)\neq\emptyset$ holds and hence $d_P\oplus d_Q$ exists. 
However, since $\tilde\Phi$ is not necessarily injective, $d_P\oplus d_Q$ is not always determined uniquely.
We choose $o=\mu(D)$ for $D\in\ker\Phi$ as the unit of addition on ${\rm Sym}^g(H)$.

Let $d_P$, $d_Q$, and $d_R$ be the elements of ${\rm Sym}^g(H)$ satisfying
\begin{align}
d_P
\oplus
d_Q
\oplus
d_R
=
o.
\label{eq:addonH}
\end{align}
This can be written by the divisors on ${\rm Pic}^0(H)$:
\begin{align*}
D_P+D_Q+D_R-3D^\ast
\equiv
0
\quad
\mbox{(mod $\mathcal{D}_l(H)$).}
\end{align*}
There exists a rational function $k\in L(3D^\ast)$ whose $3g$ zeros are $P_1,\ldots,P_g$, $Q_1,\ldots,Q_g$, $R_1,\ldots,R_g$.
Let $C$ be the curve defined by $k$: $C=(k(x,y)=0)$.
Then the zeros of $k$ are points on $C$.
Since these points are on $H$ by definition, these points are the intersection points of $H$ and $C$.
Thus (\ref{eq:addonH}) is realized by using the intersection of $H$ and $C$.

\subsection{Kernel of $\Phi$}
Hereafter, we fix $D^\ast$ as follows
\begin{align*}
D^\ast
=
\begin{cases}
\displaystyle\frac{g}{2}(P_\infty+P_\infty^\prime)&\mbox{for even $g$,}\\
\displaystyle\frac{g-1}{2}(P_\infty+P_\infty^\prime)+P_\infty&\mbox{for odd $g$.}\\
\end{cases}
\end{align*}
Put $I_n:=\{1,2,\ldots,n\}$ for $n\in\N$.
We have the following theorems concerning the kernel of the surjective canonical map $\Phi$.
\begin{theorem}\label{thm:ker}
If $g$ is an even number we have
\begin{align*}
\ker \Phi
=
\left\{
\left.
\sum_{i=1}^gP_i\ 
\right|\
\mbox{${}^\forall i\in I_g$ ${}^\exists j\in I_g$ s.t. $P_j=P_i^\prime$, $j\neq i$}
\right\}.
\end{align*}
If $g$ is an odd number we have
\begin{align*}
\ker \Phi
=
\left\{
\left.
\sum_{i=1}^{g-1}P_i+P_\infty\ 
\right|\
\mbox{${}^\forall i\in I_{g-1}$ ${}^\exists j\in I_{g-1}$ s.t. $P_j=P_i^\prime$, $j\neq i$.}
\right\}.
\end{align*}
\end{theorem}

(Proof)\quad
See \ref{app:proof1}.
\qed

If $g=1$ the surjection $\Phi$ is also injective (see theorem\ref{thm:phisurjec}).
On the other hand, for $g\geq2$, $\Phi$ is `almost' injective.
\begin{theorem}\label{thm:ker2}
For any $g$, we have
\begin{align*}
\Phi(D_P)
=
\Phi(D_Q)
\quad
\Longleftrightarrow
\quad
\mbox{$^{\exists} D_R\in\mathcal{D}_g^+(H)$ s.t. $D_P+D_{Q^\prime}=D_R+D_{R^\prime}$.}
\end{align*}
\end{theorem}

(Proof)\quad
See \ref{app:proof2}.
\qed

In particular, if $g=2$ the surjection $\Phi$ is injective except for its kernel.
\begin{corollary}\label{cor:ker3}
If $g=2$ we have
\begin{align*}
\mbox{$
\Phi(D_P)
=
\Phi(D_Q)
$ and $D_P\neq D_Q$}
\qquad
\Longleftrightarrow
\qquad
D_P, D_Q\in \ker \Phi.
\end{align*}
\end{corollary}

(Proof)\quad
Put $g=2$.
Assume
\begin{align*}
D_P+D_{Q^\prime}=R_1+R_2+R_1^\prime+R_2^\prime,
\end{align*}
where $R_1$ and $R_2$ are points on $H$.
Then we have
\begin{align*}
&\begin{cases}
D_P=R_1+R_2,\\
D_{Q}=R_1+R_2,\\
\end{cases}\quad
\begin{cases}
D_P=R_1+R_1^\prime,\\
D_{Q}=R_2+R_2^\prime,\\
\end{cases}\quad
\begin{cases}
D_P=R_1+R_2^\prime,\\
D_{Q}=R_1+R_2^\prime,\\
\end{cases}\\
&\begin{cases}
D_P=R_1^\prime+R_2,\\
D_{Q}=R_1^\prime+R_2,\\
\end{cases}\quad
\begin{cases}
D_P=R_1^\prime+R_2^\prime,\\
D_{Q}=R_1^\prime+R_2^\prime,\\
\end{cases}\quad
\begin{cases}
D_P=R_2+R_2^\prime,\\
D_{Q}=R_1+R_1^\prime.\\
\end{cases}
\end{align*}
Thus, by theorem \ref{thm:ker}, if and only if $D_P\neq D_Q$ then $D_P, D_Q\in \ker \Phi$.
\qed

\section{Discrete Toda lattice with a periodic boundary condition}
\label{sec:GRPDTL}
Let us consider the periodic discrete Toda lattice (pdTL) \cite{HTI93}, which is a map $\chi:\mathbb{C}^{2g+2}\to\mathbb{C}^{2g+2}$ given by
\begin{align*}
\chi:\ (I_1,\cdots,I_{g+1},V_1,\cdots,V_{g+1})\mapsto(\bar I_1,\cdots,\bar I_{g+1},\bar V_1,\cdots,\bar V_{g+1}),
\end{align*}
where $\bar I_i$ and $\bar V_i$ $(i=1,2,\ldots,g+1)$ are defined to be
\begin{align*}
\bar I_{i}
=
V_i
+
I_i
\frac
{1-\frac{V_1V_2\cdots V_{g+1}}{I_1I_2\cdots I_{g+1}}}
{1+\frac{V_{i-1}}{I_{i-1}}+\frac{V_{i-1}V_{i-2}}{I_{i-1}I_{i-2}}+\cdots+\frac{V_{i-1}V_{i-2}\cdots V_{i+1}}{I_{i-1}I_{i-2}\cdots I_{i+1}}},
\qquad
\bar V_i=\frac{I_{i+1}V_i}{\bar I_i}.
\end{align*}
If we assume $0<\prod_{i=1}^{g+1}V_i<\prod_{i=1}^{g+1}I_i$ these evolution equations are equivalent to the following \cite{KT02}
\begin{align}
\bar I_i+\bar V_{i-1}=I_i+V_i,
\qquad
\bar V_i\bar I_i=I_{i+1}V_i.
\label{eq:pdTL}
\end{align}
For $t=0,1,\ldots$, we use the notation
\begin{align*}
\left(I_1^t,\ldots,I_{g+1}^t,V_1^t,\ldots,V_{g+1}^t\right)
:=
\underset{t}{\underbrace{\chi\circ\chi\circ\cdots\circ\chi}}(I_1,\cdots,I_{g+1},V_1,\cdots,V_{g+1}).
\end{align*}

The Lax form of (\ref{eq:pdTL}) is given by
\begin{align}
\bar LM
=
ML,
\label{eq:pdTLLax}
\end{align}
where $L$ and $M$ are the following $(g+1)\times(g+1)$ matrices\footnote{These matrices $L$ and $M$ are for $g\geq2$. For the case of $g=1$, $L$ and $M$ are given in \ref{subsec:g1}.}
\begin{align*}
L
:=
&\left[
\begin{matrix}
I_2+V_1&1&&&(-1)^gI_1V_1/y\\
I_2V_2&I_3+V_2&1&&\\
&\ddots&\ddots&\ddots&\\
&&I_gV_g&I_{g+1}+V_g&1\\
(-1)^gy&&&I_{g+1}V_{g+1}&I_1+V_{g+1}\\
\end{matrix}
\right],
\\
M
:=
&\left[
\begin{matrix}
I_2&1&&&\\
&I_3&1&&\\
&&\ddots&\ddots&\\
&&&I_{g+1}&1\\
(-1)^gy&&&&I_1
\end{matrix}
\right]
\end{align*}
and $\bar L$ is obtained from $L$ by replacing $I_i$ and $V_i$ with $\bar I_i$ and $\bar V_i$ ($i=1,2,\ldots,g+1$), respectively.
Here $y\in\C$ is the spectral parameter.

\subsection{Conserved quantities}
Let $\Lambda$ be the set $\{1,2,\ldots,g+1\}$.
For $\Lambda$, we define the $(g+1)\times(g+1)$ triple diagonal matrix $L_\Lambda$ to be
\begin{align*}
L_\Lambda
:=
&\left[
\begin{matrix}
I_2+V_1&1&&\\
I_2V_2&\ddots&\ddots&\\
&\ddots&\ddots&1\\
&&I_{g+1}V_{g+1}&I_1+V_{g+1}\\
\end{matrix}
\right].
\end{align*}
Note that $L_\Lambda$ is the Lax matrix of the discrete Toda molecule \cite{HTI93}.
We denote the $(i,j)$-entry of $L_\Lambda$ by $\lambda_{ij}$.
For a subset $\Omega\subset\Lambda$, we define $L_\Omega:=\left(\lambda_{ij}\right)_{i,j\in\Omega}$ as well.

Let us consider the polynomial $\tilde f(x,y)$ in $x$ and $y$
\begin{align*}
&\tilde f(x,y)
:=
y\left|
x\mathbb{I}+L
\right|,
\end{align*}
where $\mathbb{I}$ is the identity matrix of degree $g+1$. 
The eigenpolynomial of $L$ can be expanded as follows
\begin{align*}
\left|
x\mathbb{I}+L
\right|
&=
y
+
\left|x\mathbb{I}+L_\Lambda\right|
-
I_1V_1\left|x\mathbb{I}+L_{\bar{\bar\Lambda}}\right|
+
\frac{c_{-1}}{y},
\end{align*}
where we put $\bar{\bar\Lambda}:=\Lambda\setminus\{1,g+1\}=\{2,3,\ldots,g\}\subset\Lambda$ and 
\begin{align}
c_{-1}
=
\prod_{i=1}^{g+1}I_iV_i.
\label{eq:cm1}
\end{align}

Note the formula
\begin{align*}
\left|x\mathbb{I}+L_\Lambda\right|
=
x^{g+1}
+
\sum_{k=1}^{g+1}\sum_{\substack{\Omega\subset\Lambda\cr |\Omega|=k}}\left| L_{\Omega}\right|x^{g+1-k},
\end{align*}
where $\Omega$ ranges over all subsets of $\Lambda$ consisting of $k$ elements. 
We find
\begin{align*}
\left|
x\mathbb{I}+L
\right|
&=
y
+
x^{g+1}
+
\sum_{k=1}^{g+1}c_{g+1-k}x^{g+1-k}
+
\frac{c_{-1}}{y},
\end{align*}
where we put
\begin{align}
&c_g
=
\sum_{i=1}^{g+1}(I_i+V_i),\label{eq:cg}\\
&c_{g+1-k}
=
\sum_{\substack{\Omega\subset\Lambda\cr |\Omega|=k}}\left| L_{\Omega}\right|-I_1V_1\sum_{\substack{\Upsilon\subset\bar{\bar\Lambda}\cr |\Upsilon|=k-2}}\left| L_{\Upsilon}\right|
\quad
\mbox{($k=2,3,\ldots,g+1$).}
\label{eq:cgother}
\end{align}
Thus we obtain 
\begin{align}
&\tilde f(x,y)
=
y^2+y\left(x^{g+1}+c_gx^g+\cdots+c_1x+c_0\right)+c_{-1}.
\label{eq:epl}
\end{align}
We may use the notation $c_i^t:=c_i\left(I_1^t,\ldots,I_{g+1}^t,V_1^t,\ldots,V_{g+1}^t\right)$ ($t=0,1,\ldots$).

Since $\left| M\right|=\prod_{i=1}^{g+1}I_i+(-1)^gy\not\equiv0$, we have
$\bar L=MLM^{-1}$  by (\ref{eq:pdTLLax}).
Hence, the eigenpolynomial of $L$ is invariant with respect to the time evolution (\ref{eq:pdTL}):
\begin{align*}
\left|
x\mathbb{I}+\bar L
\right|
&=
\left|
x\mathbb{I}+MLM^{-1}
\right|
=
\left|
x\mathbb{I}+L
\right|.
\end{align*}
Thus the coefficients $c_{-1},c_{0},\ldots,c_{g}$ are the conserved quantities of the pdTL.

We show that the conserved quantity $c_{g+1-k}$ is a homogeneous polynomial in $I_i$ and $V_j$ of degree $k$.
A subset $\Omega\subset\Lambda$ such that $|\Omega|=k$ is a union of sets of consecutive numbers:
\begin{align*}
&\Omega
=
\bigcup_{s=1}^l\Omega_{s},\qquad
\Omega_{s}
=
\left\{
\omega_{s},\omega_{s}+1,\ldots,\omega_{s}+|\Omega_s|-1
\right\},
\end{align*}
where we assume $\omega_s<\omega_t$ for $s<t$.
Therefore, the matrix $L_{\Omega}$ is block diagonal. 
This implies $\left| L_{\Omega} \right|=\prod_{s=1}^{l}\left| L_{\Omega_{s}}\right|$.
The determinant $\left| L_{\Omega_{s}}\right|$ is given by
\begin{align*}
\left| L_{\Omega_{s}}\right|
=
\sum_{m=0}^{|\Omega_{s}|}\prod_{i=m+1}^{|\Omega_{s}|}I_{i+\omega_s}\prod_{j=0}^{m-1}V_{j+\omega_s},
\end{align*}
where we assume $\prod_{i=|\Omega_{s}|+1}^{|\Omega_{s}|}I_{i+\omega_s}=\prod_{j=0}^{-1}V_{j+\omega_s}=1$.

Now assume $1,g+1\in\Omega$ for $\Omega=\bigcup_{s=1}^l\Omega_{s}\subset\Lambda$ such that $|\Omega|=k$.
Then, by definition, we find $1\in\Omega_{1}$ and $g+1\in\Omega_{l}$.
Put $\bar\Omega_{1}:=\Omega_{1}\setminus\{1\}$ and $\bar\Omega_{l}:=\Omega_{l}\setminus\{g+1\}$.
Then we have
\begin{align*}
\left| L_{\Omega_{1}}\right|
&=
\prod_{i=2}^{|\Omega_1|+1}I_{i}
+
V_1\left| L_{\bar\Omega_{1}}\right|,
\qquad
\left| L_{\Omega_{l}}\right|
=
\prod_{j=\omega_l}^{g+1}V_{j}
+
I_1\left| L_{\bar\Omega_{l}}\right|,
\end{align*}
where we use the fact $\omega_l+|\Omega_l|-1=g+1$.
Moreover, set $\tilde\Omega:=\bigcup_{s=2}^{l-1}\Omega_{s}$, $\bar\Omega:=\bar\Omega_{1}\cup\tilde\Omega\cup\bar\Omega_{l}$, $\bar\Omega_{\bot}:=\tilde\Omega\cup\bar\Omega_{l}$, and $\bar\Omega_{\top}:=\bar\Omega_{1}\cup\tilde\Omega$.
Then, for $\Omega=\bigcup_{s=1}^l\Omega_{s}$ such that $1\in\Omega_1$ and $g+1\in\Omega_l$, we have
\begin{align*}
\left| L_{\Omega} \right|
&=
\prod_{i=2}^{|\Omega_1|+1}I_{i}
\prod_{j=\omega_l}^{g+1}V_{j}
\left| L_{\tilde\Omega} \right|
+
\prod_{i=1}^{|\Omega_1|+1}I_{i}
\left| L_{\bar\Omega_{\bot}}\right|
+
\prod_{j=\omega_l}^{g+2}V_{j}
\left| L_{\bar\Omega_{\top}}\right|
+
I_1V_1\left| L_{\bar\Omega} \right|.
\end{align*}
Noting $\bar\Omega\subset\bar{\bar\Lambda}$ and $|\bar\Omega|=k-2$, we obtain
\begin{align*}
c_{g+1-k}
&=
\sum_{\substack{\Omega^\prime\subset\Lambda\cr |\Omega^\prime|=k}}\left| L_{\Omega^\prime}\right|\nn\\
&+
\sum_{\substack{\Omega\subset\Lambda\cr |\Omega|=k}}
\left(
\prod_{i=2}^{|\Omega_1|+1}I_{i}
\prod_{j=\omega_l}^{g+1}V_{j}
\left| L_{\tilde\Omega} \right|
+
\prod_{i=1}^{|\Omega_1|+1}I_{i}
\left| L_{\bar\Omega_{\bot}}\right|
+
\prod_{j=\omega_l}^{g+2}V_{j}
\left| L_{\bar\Omega_{\top}}\right|
\right),
\end{align*}
where the first sum is taken over all $\Omega^\prime\subset\Lambda$ which does not contain both $1$ and $g+1$, and the second sum is taken over all $\Omega\subset\Lambda$ which contains both $1$ and $g+1$. 
We assume $|L_\emptyset|=1$.
Thus we obtain the following.
\begin{proposition}\label{prop:subfree}
The coefficient $c_{g+1-k}$ of $x^{g+1-k}$ in $\tilde f(x,y)$ is a subtraction-free homogeneous polynomial in $I_1,\ldots,I_{g+1},V_1,\ldots,V_{g+1}$ of degree $k$ for $k=1,2,\ldots,g+1$.
\qed
\end{proposition}

\begin{example}
Put $k=g+1$. 
Then there exists no $\Omega^\prime$ which does not contain both 1 and $g+1$, and we find $\Omega=\Lambda$ with $|\Omega|=g+1$.
This implies $\Omega_1=\Omega$, $\tilde\Omega=\Omega_l=\emptyset$.
Therefore, we obtain
\begin{align*}
c_{0}
=
\prod_{i=1}^{g+1}I_{i}
+
\prod_{j=1}^{g+1}V_{j}.
\end{align*}

Put $k=g$.
Then we find $\Omega^\prime=\bar\Lambda:=\Lambda\setminus\{g+1\}=\{1,2,\ldots,g\}$ or $\Omega^\prime=\{2,\ldots,g,g+1\}$.
We also find $\Omega=\Omega_1\cup\Omega_2$, $\Omega_1=\{1,2,\ldots,n-1\}$, and $\Omega_2=\{n+1,n+2,\ldots,g+1\}$ for $n=2,3,\ldots,g$.
Noting $|\Omega_1|=n-1$ and $\omega_2=n+1$, we obtain
\begin{align}
c_{1}
&=
\sum_{m=0}^g\left(\prod_{i=m+2}^{g+1}I_i\prod_{j=1}^{m}V_j
+
\prod_{i=m+3}^{g+2}I_i\prod_{j=2}^{m+1}V_j\right)
+
\sum_{n=2}^{g+1}
\prod_{i=2}^{n}I_i
\prod_{j=n+1}^{g+1}V_j\nn\\
&\quad
+
\sum_{n=2}^g
\left(
\prod_{i=1}^{n}I_i
\sum_{m=0}^{g-n}\prod_{i=m+n+2}^{g+1}I_{i}\prod_{j=1+n}^{m+n}V_{j}
+
\prod_{j=n+1}^{g+2}V_j
\sum_{m=0}^{n-2}\prod_{i=m+3}^{n}I_{i}\prod_{j=2}^{m+1}V_{j}
\right).
\label{eq:c1coef}
\end{align}
\end{example}

\subsection{Spectral curves}
Let us consider the affine curves on $\C^2_{(x,y)}$ and $\C^2_{(u,v)}$, respectively
\begin{align*}
&\tilde\gamma_0
:=
\left(
\tilde f(x,y)=0
\right),\\
&\tilde\gamma_\infty
:=
\left(
v^2+v\left(1+c_gu+\cdots+c_1u^{g}+c_0u^{g+1}\right)+c_{-1}u^{2g+2}=0
\right).
\end{align*}
These affine curves are non-singular under certain conditions for $c_i$'s.
Let the projections on the curves be
\begin{align*}
\pi_0:\tilde\gamma_0\to\C_x;\ (a,b)\mapsto a,\quad
\pi_\infty:\tilde\gamma_\infty\to\C_u;\ (\alpha,\beta)\mapsto \alpha.
\end{align*}
By gluing $\tilde\gamma_0$ with $\tilde\gamma_\infty$ in terms of the bi-holomorphic map $\tilde\gamma_0\setminus\pi_0^{-1}(0)\to\tilde\gamma_\infty\setminus\pi_\infty^{-1}(0)$;
\begin{align*}
(x,y)\mapsto(u,v)=
\left(\frac{1}{x},
\frac{y}{x^{g+1}}\right),
\end{align*}
we obtain the hyperelliptic curve $\tilde\gamma=\tilde\gamma_0\cup\tilde\gamma_\infty$ of genus $g$.

Substitute $x=1/t$ into $\tilde f(x,y)=0$.
We then find
\begin{align*}
\left(t^{g+1}y\right)^2
+
\left(t^{g+1}y\right)
\left(1+c_g t+\cdots+c_1 t^g+c_0 t^{g+1}\right)
+
c_{-1}t^{2g+2}
=
0.
\end{align*}
For sufficiently small $t$ this equation reduces to $\left(t^{g+1}y\right)^2+\left(t^{g+1}y\right)\sim0$.
Hence, we obtain $(x,y)\sim\left({1}/{t},0\right),\left({1}/{t}, -{1}/{t^{g+1}}\right)$, or equivalently $(u,v)\sim\left(t,0\right),\left(t, -1\right)$.
Thus the curve $\tilde\gamma$ can be expressed as
\begin{align*}
&\tilde\gamma
=
\tilde\gamma_0\cup
\left\{
\tilde P_\infty,\tilde P_\infty^\prime
\right\},
\end{align*}
where $\tilde P_\infty:=(0,0), \tilde P_\infty^\prime:=(0,-1)\in\C_{(u,v)}$ are the points at infinity.

Now we consider the rational map $\rho:\C^2_{(x,y)}\to\C^2_{(u,v)}$;
\begin{align*}
\rho: (x,y)
\mapsto
(u,v)
&=
\left(
x,
y-\frac{c_{-1}}{y}
\right).
\end{align*}
By applying $\rho$ to points on $\tilde\gamma_0$, we obtain the affine curve $\gamma_0$ on $\C_{(u,v)}^2$:
\begin{align*}
&\gamma_0
:=
\left(
f(u,v)=0
\right)
=
\left\{
(u,v)
=
\rho(x,y)\ |\ 
\tilde f(x,y)=0
\right\},\\
&f(u,v):=
v^2
-
\left(
u^{g+1}+c_gu^g+\cdots+c_1u+c_0
\right)^2
+
4c_{-1}.
\end{align*}
As in the same manner in section \ref{sec:AOH}, by gluing $\gamma_0$ with the affine curve
\begin{align*}
&\gamma_\infty
:=
\left(
y^2
-
\left(
1+c_gx+\cdots+c_1x^g+c_0x^{g+1}
\right)^2
+
4c_{-1}x^{2g+2}=0
\right),
\end{align*}
we obtain the canonical hyperelliptic curve $\gamma=\gamma_0\cup\gamma_\infty$ of genus $g$.
Remember that the curve $\gamma$ can be given by
\begin{align*}
&\gamma
=
\gamma_0\cup
\left\{
P_\infty,P_\infty^\prime
\right\},
\end{align*}
where $P_\infty=(0,1),\ P_\infty^\prime=(0,-1)\in\C_{(x,y)}$ are the points at infinity.

Let the solution to the quadratic equation $\tilde f(1/t,y)=0$ in $y$ be $y_{+}$ and $y_{-}$ which tend to $0$ and $-1/t^{g+1}$ for sufficiently small $t$, respectively. 
Apply the rational map $\rho$ to the points $(1/t,y_{+}),(1/t,y_{-})\in\tilde\gamma_0$.
Then, for sufficiently small $t$, we find
\begin{eqnarray*}
&\rho\left(\frac{1}{t},y_{+}\right)
\sim
\left(\frac{1}{t},\frac{1}{t^{g+1}}\right),
\qquad
&\rho\left(\frac{1}{t},y_{-}\right)
\sim
\left(\frac{1}{t},-\frac{1}{t^{g+1}}\right).
\end{eqnarray*}
Therefore, the map $\rho$ is naturally extended to on the hyperelliptic curve $\tilde\gamma$:
\begin{align*}
\rho(\tilde P_\infty)=P_\infty,\qquad
\rho(\tilde P_\infty^\prime)=P_\infty^\prime.
\end{align*}
We also denote the extension by $\rho$.
The hyperelliptic curves $\tilde\gamma$ and $\gamma$ are called the spectral curves of the pdTL.

\subsection{Eigenvector maps}
Let the phase space of the pdTL be 
\begin{align*}
\mathcal{U}:=\left\{(I_1,\ldots,I_{g+1},V_1,\ldots,V_{g+1})\ |\ 0<\prod_{i=1}^{g+1}V_i<\prod_{i=1}^{g+1}I_i\right\}.
\end{align*}
Also let the moduli space of $\gamma$ be $\mathcal{C}:=\left\{(c_{-1},c_0,\ldots,c_g)\right\}$.
Consider the map
\begin{align*}
\psi:\mathcal{U}\to\mathcal{C};\ 
(I_1,\ldots,I_{g+1},V_1,\ldots,V_{g+1})\mapsto (c_{-1},c_0,\ldots,c_g)
\end{align*}
defined by (\ref{eq:cm1} - \ref{eq:cgother}).
We define the isolevel set $\mathcal{U}_c$ of the pdTL to be 
\begin{align*}
\mathcal{U}_c:=\psi^{-1}(c_{-1},c_0,\ldots,c_g)\subset\mathcal{U}.
\end{align*}
The isolevel set $\mathcal{U}_c$ is isomorphic to the affine part of the Jacobian $J(\gamma)$ of $\gamma$, and the time evolution (\ref{eq:pdTL}) is linearized on it \cite{AvM80,KT02,Iwao08}.

Let $\varphi(x,y)=\left(\varphi_1,\cdots,\varphi_g,-\varphi_{g+1}\right)^T$ be the eigenvector of the Lax matrix $L$ associated with the eigenvalue $x$.
Consider the eigenvalue equation of $L$
\begin{align}
\left(
x\mathbb{I}+L
\right)
\varphi(x,y)
=
0.
\label{eq:sse}
\end{align}
Let the $(i,j)$-entry of the matrix $L$ be $l_{ij}$.
By applying the Cramer formula, each element of $\varphi(x,y)$ is explicitly given by
\begin{align*}
\varphi_i(x,y)
=
\left|
\begin{matrix}
&l_{11}+x&\cdots&l_{1,i-1}&l_{1,g+1}&l_{1,i+1}&\cdots&l_{1g}\\
&\vdots&&\vdots&\vdots&\vdots&&\vdots\\
&l_{g1}&\cdots&l_{g,i-1}&l_{g,g+1}&l_{g,i+1}&\cdots&l_{gg}+x\\
\end{matrix}
\right|
\end{align*}
for $i=1,2,\ldots,g$ and
\begin{align*}
\varphi_{g+1}(x)
=
\left|x\mathbb{I}+L_{\bar\Lambda}\right|.
\end{align*}
Thus we see that $\varphi_{g+1}(x)$ is the eigenpolynomial of $L_{\bar\Lambda}$.

We introduce another expression of $\varphi_{g+1}(x)$.
Expand the eigenpolynomial $\left|x\mathbb{I}+L\right|$ with respect to the last row:
\begin{align}
\left|x\mathbb{I}+L\right|
=
y\varphi_1(x,y)
-
I_{g+1}V_{g+1}\varphi_{g}(x,y)+
\left(x+I_1+V_{g+1}\right)\varphi_{g+1}(x).
\label{eq:eplexp}
\end{align}
To show several properties of the spectral curves, we will set the variables $x, y$ and parameters $I_1, I_{g+1}, V_1, V_{g+1}$ specially. 
For this purpose, we denote $\left|x\mathbb{I}+L\right|$ by $\theta(x,y;I_1,V_1,I_{g+1},V_{g+1})$.
By setting $I_{1}=V_{g+1}=0$ in (\ref{eq:eplexp}), we obtain an expression of $\varphi_{g+1}(x)$
\begin{align*}
\varphi_{g+1}(x)
=
\frac{\theta(x,y;0,I_{g+1},V_1,0)-y}{x}.
\end{align*}

Now we define the eigenvector map \cite{vMM79,Iwao08}.
Each equation in (\ref{eq:sse}) is a three-term relation among $\varphi_{i-1}$, $\varphi_i$, and $\varphi_{i+1}$. 
Therefore, if a point $(x,y)$ satisfies $\varphi_1(x,y)=\varphi_2(x,y)=0$ then $\varphi_i(x,y)=0$ holds for all $i=1,2,\ldots,g+1$. 
We can easily see that there exactly exist $g$ points $(x,y)$ satisfying $\varphi_1(x,y)=\varphi_2(x,y)=0$, counting multiplicities.
Denote these points by $\tilde P_1,\tilde P_2,\cdots,\tilde P_g$.
By eliminating $y$ from $\varphi_1(x,y)=0$ and $\varphi_2(x,y)=0$, we obtain $\varphi_{g+1}(x)=\left|x\mathbb{I}+L_{\bar \Lambda}\right|=0$.
Thus the $x$-component of $\tilde P_i$ ($i=1,2,\ldots,g$) is the eigenvalue of $L_{\bar\Lambda}$.
Moreover, we see that all $\tilde P_i$'s are on the spectral curve $\tilde \gamma$.
Because the $(g+1)$-th equation
\begin{align*}
(-1)^gy\varphi_1(x,y)
+
I_{g+1}V_{g+1}\varphi_g(x,y)
-
\left(I_1+V_{g+1}+x\right)\varphi_{g+1}(x)
=
0
\end{align*}
in (\ref{eq:sse}) is equivalent to $\tilde f(x,y)=0$, the defining equation of $\tilde \gamma$.
We choose $D_P=P_1+\cdots+P_g$, where $P_i:=\rho(\tilde P_i)$, as a representative of ${\rm Pic}^g\mkern2mu(\gamma):=\mathcal{D}_g(\gamma)/\mathcal{D}_l(\gamma)$.
We define the eigenvector map
\begin{align*}
\phi:\mathcal{U}_c\to{\rm Pic}^g\mkern2mu(\gamma);\ 
U\mapsto\phi(U)\equiv D_P
\quad
\mbox{(mod $\mathcal{D}_l(\gamma)$)}.
\end{align*}

Let the subset $\mathcal{D}$ of $\mathcal{D}_g^+(\gamma)$ be
\begin{align*}
\mathcal{D}
:=
\left\{
\left.D_P\in\mathcal{D}_g^+(\gamma)\ \right | \ 
\phi(U)\equiv D_P
\
\mbox{(mod $\mathcal{D}_l(\gamma)$) for $U\in\mathcal{U}_c$}
\right\}.
\end{align*}
\begin{theorem}\label{th:phig}
The reduced map $\Phi|{\mathcal{D}}:\mathcal{D}\to{\rm Pic}^0\mkern2mu(\gamma)$ is injective.
\end{theorem}

(Proof)\quad
If $\Phi(D_{P})\equiv\Phi(D_{Q})$ then, by theorem \ref{thm:ker2}, we have
\begin{align}
D_{P}
+
D_{Q^\prime}
=
D_R+D_{R^\prime}
\label{eq:pqprim}
\end{align}
for some $D_R\in\mathcal{D}_g^+(\gamma)$.
It immediately follows
\begin{align}
D_{P^\prime}
+
D_{Q}
=
D_R+D_{R^\prime}.
\label{eq:pqprim2}
\end{align}
Subtract (\ref{eq:pqprim2}) from (\ref{eq:pqprim}). 
We then obtain
\begin{align}
D_{P}
-
D_{P^\prime}
+
D_{Q^\prime}
-
D_{Q}
=
0.
\label{eq:dpmdq}
\end{align}

Let $D_{P}=\sum_{i=1}^gP_i$ be the representative of ${\rm Pic}^g\mkern2mu(\gamma)$ such that $\phi(U)\equiv D_{P}$ (mod $\mathcal{D}_l(\gamma)$) for $U\in\mathcal{U}_c$.
Let $P_1,\cdots,P_k$ be the bifurcation points and $P_{k+1},\cdots,P_g$ ordinary points.
Then, by construction of $D_{P}$, we have
\begin{align}
&P_i
=
P_i^\prime
\quad
i=1,\ldots,k,\label{eq:p1}\\
&\left\{P_{k+1},\cdots,P_g\right\}
\cap
\left\{P_{k+1}^\prime,\cdots,P_g^\prime\right\}
=
\emptyset.\label{eq:p2}
\end{align}
Also let $D_{Q}=\sum_{i=1}^gQ_i$ be the representative of ${\rm Pic}^g\mkern2mu(\gamma)$ such that $\phi(V)\equiv D_{Q}$ (mod $\mathcal{D}_l(\gamma)$) for $V\in\mathcal{U}_c$.
Let $Q_1,\cdots,Q_l$ be the bifurcation points and $Q_{l+1},\cdots,Q_g$ ordinary points.
Then we have
\begin{align}
&Q_i
=
Q_i^\prime
\quad
i=1,\ldots,l,\label{eq:q1}\\
&\left\{Q_{l+1},\cdots,Q_g\right\}
\cap
\left\{Q_{l+1}^\prime,\cdots,Q_g^\prime\right\}
=
\emptyset.\label{eq:q2}
\end{align}
By (\ref{eq:dpmdq}), (\ref{eq:p1}), and (\ref{eq:q1}), we find
\begin{align*}
D_{P}
-
D_{P^\prime}
+
D_{Q^\prime}
-
D_{Q}
=
\sum_{i=k+1}^gP_i
-
\sum_{i=k+1}^gP_i^\prime
+
\sum_{i=l+1}^gQ_i^\prime
-
\sum_{i=l+1}^gQ_i
=
0.
\end{align*}
By (\ref{eq:p2}) and (\ref{eq:q2}), we further obtain
\begin{align*}
\sum_{i=k+1}^gP_i
=
\sum_{i=l+1}^gQ_i,\qquad
\sum_{i=k+1}^gP_i^\prime
=
\sum_{i=l+1}^gQ_i^\prime.
\end{align*}
This implies $k=l$ and $D_{P}=D_{Q}$.
\qed

\subsection{Time evolution}
In the linearization of time evolution of the pdTL on the Jacobian $J(\gamma)$, we choose the element $D_P\equiv\phi(U)$ (mod $\mathcal{D}_l(\gamma)$) of $\mathcal{D}_g^+(\gamma)$ as the representative of an element $U$ of the isospectral set $\mathcal{U}_c$.
Theorem \ref{th:phig} shows that the subset $\mathcal{D}$ of $\mathcal{D}_g^+(\gamma)$ consisting of such representatives is injectively mapped into ${\rm Pic}^0(\gamma)$ by the canonical map $\Phi$.
Therefore, the time evolution of pdTL is also linearized on ${\rm Pic}^0(\gamma)$ through the injection
\begin{align*}
\mathcal{U}_c\to{\rm Pic}^0(\gamma);\  
U\mapsto \left(\Phi\circ\phi\right)(U)
=
\Phi(D_P).
\end{align*}

It should be noted the following theorem concerning the linearization of time evolution of the pdTL shown by Iwao \cite{Iwao08,Iwao10-2}.
\begin{theorem}(Proposition 2.16 in \cite{Iwao08} with $N=g+1$ and $M=1$)
\label{thm:iwao}
Let $D$ be the divisor $D=A-P_\infty^\prime$, where $A=\left(0,\prod_{i=1}^{g+1}V_i-\prod_{i=1}^{g+1}I_i\right)$.
Then the following diagram is commutative
\begin{align*}
\begin{CD}
\mathcal{U}_c @> \phi >> {\rm Pic}^g(\gamma)\\
@V (\ref{eq:pdTL}) VV @VV +D V\\
\mathcal{U}_c @>> \phi > {\rm Pic}^g(\gamma).\\
\end{CD}
\end{align*}
\qed
\end{theorem}

By theorem \ref{thm:iwao}, there exists a rational function $h$ on $\gamma$ such that $D_{\bar P}=D_{P}+D+(h)$.
It immediately follows
\begin{align*}
D_{\bar P}-D^\ast
\equiv
D_{P}-D^\ast
+
D
\quad
\mbox{(mod $\mathcal{D}_l(\gamma))$.}
\end{align*}
Note that this is an addition formula on ${\rm Pic}^0(\gamma)$.

Consider the divisor
\begin{align*}
T
:=
D+D^\ast
=
\begin{cases}
\displaystyle A+\frac{g-2}{2}\left(P_\infty+P_\infty^\prime\right)+P_\infty&\mbox{for even $g$}\\
\displaystyle A+\frac{g-3}{2}\left(P_\infty+P_\infty^\prime\right)+2P_\infty&\mbox{for odd $g$.}\\
\end{cases}
\end{align*}
We observe that $T\in\mathcal{D}_g^+(\gamma)$ for $g\geq2$, and hence we have
\begin{align*}
\Phi(T)
\equiv
D
\quad
\mbox{(mod $\mathcal{D}_l(\gamma)$).}
\end{align*}
Since $\mu:\mathcal{D}_g^+(\gamma)\to{\rm Sym}^g(\gamma)$ is bijective, we obtain the following theorem.

\begin{theorem}\label{thm:linearization}
Let $\tau:=\mu(T)\in{\rm Sym}^g(\gamma)$ for $g\geq2$.
Then the following diagram is commutative\footnote{For the case of $g=1$, see \ref{subsec:g1}.}
\begin{align*}
\begin{CD}
\mathcal{U}_c @> \mu\circ\phi >> {\rm Sym}^g(\gamma)\\
@V (\ref{eq:pdTL}) VV @VV \oplus\tau V\\
\mathcal{U}_c @>> \mu\circ\phi > {\rm Sym}^g(\gamma).\\
\end{CD}
\end{align*}
\qed
\end{theorem}

Thus the following addition formula on ${\rm Sym}^g(\gamma)$ is equivalent to the time evolution (\ref{eq:pdTL}) of the pdTL
\begin{align}
d_{\bar P}
=
d_{P}
\oplus
\tau.
\label{eq:tepdtl}
\end{align}

\section{A geometric realization of pdTL}
\label{sec:grpdtl}
In this section, we realize the time evolution (\ref{eq:tepdtl}) of the pdTL on ${\rm Sym}^g(\gamma)$ by using curve intersections. 

\subsection{Linear spaces of rational functions}
The addition formula (\ref{eq:tepdtl}) on ${\rm Sym}^g(\gamma)$ reduces to
\begin{align}
-d_{\bar P}
\oplus
d_{P}
\oplus
\tau
=
o,
\label{eq:tepdtl2}
\end{align}
where $-d_{\bar P}$ is the inverse of $d_{\bar P}$ with respect to $\oplus$.
This can be written by the divisors on ${\rm Pic}^0(\gamma)$:
\begin{align}
D_Q
+
D_P
+
A
-
P_\infty^\prime
-
2D^\ast
\equiv
0
\quad
\mbox{(mod $\mathcal{D}_l(\gamma)$)},
\label{eq:lindiv}
\end{align}
where we put $D_Q:=\mu^{-1}(-d_{\bar P})$.
This implies that there exists a rational function $h$ on $\gamma$ such that $(h)=D_Q+D_P+A-P_\infty^\prime-2D^\ast$ and $h\in L(P_\infty^\prime+2D^\ast)$. 
The curve given by $h$ passes through the points $A$, $P_1,\ldots,P_g$, $Q_1,\ldots,Q_g$ on $\gamma$.

In order to obtain the curves given by rational functions in $L(P_\infty^\prime+2D^\ast)$, we first establish the linear space $L(P_\infty^\prime+2D^\ast)$.
For a while, we assume that $g$ is an even number.
We then have $L(P_\infty^\prime+2D^\ast)=L\left(gP_\infty+(g+1)P_\infty^\prime\right)$.
The principal divisors of the coordinate functions $x$ and $y$ are 
\begin{align*}
&(x)
=
P_0+P_0^\prime
-
P_\infty-P_\infty^\prime\quad
\mbox{and}\quad
(y)
=
\sum_{i=1}^{2g+2}P_{\alpha_i}
-
(g+1)\left(P_\infty+P_\infty^\prime\right),
\end{align*}
respectively, where $P_0$ is the point on $\gamma$ whose $x$-component is 0 and $P_{\alpha_i}$ is the point on $\gamma$ whose $x$-component solves $f(x,0)=0$.
We can easily see $\left\langle1,x,\cdots,x^g\right\rangle\subset L(P_\infty^\prime+2D^\ast)$ and $y\not\in L(P_\infty^\prime+2D^\ast)$.

Let the local parameter at $P_\infty$ be $t$.
Substitute $x=1/t$ into $f(x,y)=0$. 
We then find $y=t^{-g-1}\left(1+c_g t+c_{g-1}t^2+\cdots+c_1t^g+c_0t^{g+1}\right)+o(1)$.
This implies
\begin{align*}
x^{g+1}-y
=
c_g t^{-g}+c_{g-1}t^{-g+1}+o(t^{-g+1}).
\end{align*}
Therefore, the rational function $x^{g+1}-y$ has a pole at $P_\infty$ whose order is smaller than or equal to $g$.
Also let the local parameter at $P_\infty^\prime$ be $s$.
Then we also find
\begin{align*}
x^{g+1}-y
=
2s^{-g-1}-c_g s^{-g}-c_{g-1}s^{-g+1}+o(s^{-g+1}).
\end{align*}
Thus the rational function $x^{g+1}-y$ has a pole at $P_\infty^\prime$ whose order is exactly $g+1$.
Therefore, we find $x^{g+1}-y\in L(P_\infty^\prime+2D^\ast)$.
By the Riemann-Roch theorem, we have $\dim L(P_\infty^\prime+2D^\ast)=g+2$.
Thus we obtain $L(P_\infty^\prime+2D^\ast)=\left\langle1,x,\cdots,x^g,x^{g+1}-y\right\rangle$.

In a similar manner, we also obtain $L(P_\infty^\prime+2D^\ast)=\left\langle1,x,\cdots,x^{g},x^{g+1}+y\right\rangle$ for an odd number $g$.

Next consider the addition formula on ${\rm Sym}^g(\gamma)$
\begin{align*}
-d_{\bar P}
\oplus
d_{\bar P}
=
o.
\end{align*}
This can be written by the divisors on ${\rm Pic}^0(\gamma)$:
\begin{align}
D_Q
+
D_{\bar P}
-
2D^\ast
\equiv
0
\quad
\mbox{(mod $\mathcal{D}_l(\gamma)$)}.
\label{eq:lindiv2}
\end{align}
Hence we consider the linear space $L(2D^\ast)$.
For an odd number $g$, we find 
\begin{align*}
L(2D^\ast)=\left\langle1,x,\cdots,x^{g-1},c_gx^g+x^{g+1}+y\right\rangle.
\end{align*}
Thus we obtain the following proposition.
\begin{proposition}
\label{prop:LP2D}
For any $g$, we have
\begin{align*}
L(P_\infty^\prime+2D^\ast)
=
\left\langle
1,x,\cdots,x^g,x^{g+1}-(-1)^{g}y
\right\rangle.
\end{align*}
Moreover, for any odd number $g$, we have
\begin{align*}
L(2D^\ast)
&=
\left\langle
1,x,\cdots,x^{g-1},c_gx^g+x^{g+1}+y
\right\rangle.
\end{align*}
\qed
\end{proposition}

\subsection{A curve passing through given points}\label{sec:acptgp}
Now we construct a curve given by a rational function in $L(P_\infty^\prime+2D^\ast)$ and passing through $g+1$ given points on $\gamma$.
We see from (\ref{eq:lindiv}) that there exists a rational function $h\in L(P_\infty^\prime+2D^\ast)$ such that $(h)=D_Q+D_P+A-P_\infty^\prime-2D^\ast$.
By proposition \ref{prop:LP2D}, we can assume that $h$ has the form $h(x,y)=\beta_0+\beta_1x+\cdots+\beta_gx^g+x^{g+1}-(-1)^{g}y$, where $\beta_0,\beta_1\ldots,\beta_g\in\C$.
By definition of $h$, we have $h(A)=0$. 
Hence, we find $\beta_0=(-1)^g\left(\prod_{j=1}^{g+1}V_j-\prod_{i=1}^{g+1}I_i\right)$.

We show an explicit formula which gives $\beta_1,\beta_2,\ldots,\beta_g$ by using the coefficients $c_1,c_2,\ldots,c_g$ of $\gamma$.
Introduce the involutions $s_j$ and $t_k$ ($j,k=1,2,\ldots,g+1$) on the phase space $\mathcal U$ of the pdTL:
\begin{align*}
&s_j:
\left(I_1,\ldots,I_j,\ldots,V_{g+1}\right)
\mapsto
\left(I_1,\ldots,-I_j,\ldots,V_{g+1}\right),\\
&t_k:
\left(I_1,\ldots,V_k,\ldots,V_{g+1}\right)
\mapsto
\left(I_1,\ldots,-V_k,\ldots,V_{g+1}\right).
\end{align*}
These involutions act naturally on the moduli space $\mathcal{C}$ of $\gamma$:
\begin{align*}
&s_j(c_{i})
:=
c_i(s_j^{-1}(I_1,\ldots,I_j,\ldots,V_{g+1}))
=
c_i(I_1,\ldots,-I_j,\ldots,V_{g+1}),\\
&t_k(c_{i})
:=
c_i(t_k^{-1}(I_1,\ldots,V_k,\ldots,V_{g+1}))
=
c_i(I_1,\ldots,-V_k,\ldots,V_{g+1})
\end{align*}
for $i=-1,0,\ldots,g$ and $j,k=1,2,\ldots,g+1$.
We then find
\begin{align}
\beta_0
=
\begin{cases}
s_1(c_0)<0&\mbox{for even $g$,}\\
t_{g+1}(c_0)>0&\mbox{for odd $g$,}\\
\end{cases}
\label{eq:betaczero}
\end{align}
where we use the assumption $0<\prod_{j=1}^{g+1}V_j<\prod_{i=1}^{g+1}I_i$.

Let $P_1,P_2,\ldots,P_g$ be the points on $\gamma$ given by the eigenvector map: $\sum_{i=1}^gP_i\equiv\left(\rho\circ\phi\right)(U)$ $(\mbox{mod $\mathcal{D}_l(\gamma)$})$ for $U\in\mathcal{U}_c$.
Assume that these points are in generic position (The generic condition for $P_i=(p_i,q_i)$ is $p_i\neq0$,$\ p_i\neq p_j$ for $i\neq j\in\{1,2,\ldots,g\}$).
We have the following theorem.
\begin{theorem}\label{thm:kappa1}
Let $h_1$ be a rational function in $L(P_\infty^\prime+2D^\ast)$.
Then the curve $\kappa_1=\left(h_1(u,v)=0\right)$ passing through the points $A=\left(0,\prod_{i=1}^{g+1}V_i-\prod_{i=1}^{g+1}I_i\right)$ and $P_1,P_2,\cdots,P_g$ is given by the formula
\begin{align}
h_1(u,v)
=
\begin{cases}
\DIS
\sum_{i=0}^gs_1(c_i)u^i
+
u^{g+1}
-
v
&\mbox{for even $g$},\\
\DIS
\sum_{i=0}^gt_{g+1}(c_i)u^i
+
u^{g+1}
+
v&\mbox{for odd $g$}.\\
\end{cases}\label{eq:rfk1}
\end{align}
\end{theorem}

(Proof)\quad
Remember that $|x\mathbb{I}+L|$ is denoted by $\theta(x,y;I_1,I_{g+1},V_1,V_{g+1})$.
By replacing $y$ and $I_1$ in $\theta(x,y;I_1,I_{g+1},V_1,V_{g+1})$ with $-y$ and $-I_1$ respectively,  we obtain (see (\ref{eq:epl}))
\begin{align*}
&\theta(x,-y;-I_1,I_{g+1},V_1,V_{g+1})
=
-y+\frac{c_{-1}}{y}
+
x^{g+1}
+
\sum_{i=0}^gs_1(c_i)x^i.
\end{align*}
Similarly, by replacing $V_{g+1}$ in $\theta(x,y;I_1,I_{g+1},V_1,V_{g+1})$ with $-V_{g+1}$, we obtain
\begin{align*}
&\theta(x,y;I_1,I_{g+1},V_1,-V_{g+1})
=
y-\frac{c_{-1}}{y}
+
x^{g+1}
+
\sum_{i=0}^gt_{g+1}(c_i)x^i.
\end{align*}
Here we use the fact $ s_1(c_{-1})= t_{g+1}(c_{-1})=-c_{-1}$.
By applying the rational map $\rho$ to these rational functions, we obtain $h_1$ defined by (\ref{eq:rfk1}).
Clearly, the rational function $h_1$ is in  $L(P_\infty^\prime+2D^\ast)$ (see proposition \ref{prop:LP2D}).

Note that the curve $\kappa_1=\left(h_1(u,v)=0\right)$ passes through the point $A\in\gamma$ because of the fact (\ref{eq:betaczero}).
Let $\tilde P_i=(\tilde p_i,\tilde q_i)$ ($i=1,2,\ldots,g$) be the points on $\tilde\gamma$ given by the eigenvector map.
Then we have $\varphi_1(\tilde p_i,\tilde q_i)=\cdots=\varphi_{g}(\tilde p_i,\tilde q_i)=0$ and $\varphi_{g+1}(\tilde p_i)=0$.
Moreover, we have (see (\ref{eq:eplexp}))
\begin{align*}
&\theta(x,-y;-I_1,I_{g+1},V_1,V_{g+1})\\
&\qquad\qquad=
-y\varphi_1(x,-y)
-
I_{g+1}V_{g+1}\varphi_{g}(x,-y)
+
\left(x-I_1+V_{g+1}\right)\varphi_{g+1}(x),\\
&\theta(x,y;I_1,I_{g+1},V_1,-V_{g+1})\\
&\qquad\qquad=
y\varphi_1(x,y)
+
I_{g+1}V_{g+1}\varphi_{g}(x,y)
+
\left(x+I_1-V_{g+1}\right)\varphi_{g+1}(x).
\end{align*}
Therefore, we find that 
\begin{align*}
\theta(\tilde p_i,\tilde q_i;-I_1,I_{g+1},V_1,V_{g+1})=0\quad
\mbox{and}
\quad
\theta(\tilde p_i,\tilde q_i;I_1,I_{g+1},V_1,-V_{g+1})=0
\end{align*}
hold for all $i$.
Hence, we finally show that $\kappa_1$ passes through $P_i=\rho(\tilde P_i)$ ($i=1,2,\ldots,g$).
If the points $P_1,P_2,\ldots,P_g$ are in generic position then the curve $\kappa_1$ is uniquely determined.
\qed

By construction, $\kappa_{1}$ also passes through the points $Q_1,Q_2,\ldots,Q_g$ satisfying (\ref{eq:lindiv}).
Thus the points $P_1,P_2,\ldots,P_g$, $Q_1,Q_2,\ldots,Q_g$, $A$ satisfying (\ref{eq:lindiv}) are on both $\gamma$ and $\kappa_1$.
Therefore, the addition formula (\ref{eq:tepdtl2}) can be realized by using the intersection of $\gamma$ and $\kappa_1$.
In order to realize the time evolution of the pdTL by using curve intersections, we have only to give the inverse $d_{\bar P}$ of $-d_{\bar P}=\mu(D_Q)$ in terms of a curve intersection.

\subsection{Inverse elements}
\label{sec:invelntp}
The inverse $d_{\bar P}\in{\rm Sym}^g(\gamma)$ of $-d_{\bar P}=\mu(D_Q)\in{\rm Sym}^g(\gamma)$ is given by the following theorem.
\begin{theorem}\label{thm:kappa2}
Let $Q_1,Q_2,\ldots,Q_g$ be the points on $\gamma$ satisfying (\ref{eq:lindiv}) and (\ref{eq:lindiv2}) (see theorem \ref{thm:kappa1}).
Also let $\bar P_1,\bar P_2,\ldots,\bar P_g$ be the points on $\gamma$ satisfying (\ref{eq:lindiv2}).
If $g$ is an even number we have $\bar P_i=Q_i^\prime$ for $i=1,2,\ldots,g$.
 
If $g$ is an odd number then $\bar P_1,\bar P_2,\ldots,\bar P_g$ are the intersection points of $\gamma$ and the curve $\kappa_{2}=\left(h_2(u,v)=0\right)$ given by
\begin{align}
h_2(u,v)
&=
\sum_{i=0}^{g-1}\left( s_1(c_{i})+2I_1\acute c_{i+1}\right)u^i
+
c_gu^g
+
u^{g+1}
+
v,
\label{eq:kappa2k}
\end{align}
where $\acute c_i$ is obtained from $c_i$ by setting $I_{1}=V_{1}=0$ for $i=1,2,\ldots,g$.
\end{theorem}

(Proof)\quad
If $g$ is an even number we can easily see $\bar P_i=Q_i^\prime$ for $i=1,2,\ldots,g$.
This implies the first part.

Next assume that $g$ is an odd number.
Note that the eigenpolynomial of $L$ is invariant with respect to the time evolution.
We then have
\begin{align*}
\theta(x,y;I_1,I_{g+1},V_1,V_{g+1})
&=
y\bar\varphi_1(x,y)
-
I_{1}V_{g+1}\bar\varphi_g(x,y)
+
(x+I_1+V_{1})\bar\varphi_{g+1}(x).
\end{align*}
Here $\bar\varphi_i$ is the $i$-th element of the eigenvector $\bar\varphi$ of $\bar L$ for $i=1,2,\ldots,g+1$.
By setting $I_1=V_1=0$, we obtain
\begin{align*}
\bar\varphi_{g+1}(x)
&=
\frac{
\theta(x,y;0,I_{g+1},0,V_{g+1})-y}{x}
=
x^g
+
\sum_{i=0}^{g-1}\acute c_{i+1} x^{i},
\end{align*}
where $\acute c_i$ is obtained from $c_i$ by setting $I_{1}=V_{1}=0$.
Note that the $x$-component $\bar p_i$ of $\bar P_i$ solves $\bar\varphi_{g+1}(x)=0$ by definition.
Therefore, we have
\begin{align}
\acute c_{i+1}
=
(-1)^{g-i}
\sum_{\substack{\Omega\subset\bar\Lambda\\|\Omega|=g-i}}\prod_{j\in\Omega}\bar p_j,
\label{eq:acuteci}
\end{align}
where $\bar\Lambda=\{1,2,\ldots,g\}$.

From (\ref{eq:lindiv2}), there exists a rational function $h_2\in L(2D^\ast)$ such that
\begin{align}
(h_2)
=
D_Q
+
D_{\bar P}
-
2D^\ast.
\label{eq:heq2}
\end{align}
Since $h_2\in L(2D^\ast)$, it can be written by $h_2(x,y)=\delta_0+\delta_1x+\cdots+\delta_{g-1}x^{g-1}+c_gx^g+x^{g+1}+y$ for $\delta_0,\delta_1,\ldots,\delta_{g-1}\in\C$ (see proposition \ref{prop:LP2D}).
The coefficients are determined by solving the system of linear equations
\begin{align}
h_2(\bar P_i)=0
\qquad
i=1,2,\ldots,g.
\label{eq:sheq2}
\end{align}
By using the Cramer formula, we find that if and only if the condition 
\begin{align*}
\bar p_i\neq \bar p_j
\quad
\mbox{for $i,j=1,2,\ldots,g$, $i\neq j$}
\end{align*}
holds then the following $\delta_i$'s $(i=0,2,\ldots,g-1)$ solve the equation (\ref{eq:sheq2})
\begin{align*}
\delta_i
&=
 s_1(c_i)
+
(-1)^{g-i}(c_g- s_1(c_g))
\sum_{\substack{\Omega\subset\bar\Lambda\\|\Omega|=g-i}}\prod_{j\in\Omega}\bar p_j
=
 s_1(c_i)
+
2I_1\acute c_{i+1}.
\end{align*}
Here we use the fact $c_g- s_1(c_g)=2I_1$ and (\ref{eq:acuteci}).

Thus we see that the rational function $h_2$ satisfying (\ref{eq:heq2}) is given by (\ref{eq:kappa2k}).
Therefore, the curve $\kappa_2=\left(h_2(u,v)=0\right)$ passes through the points $Q_1,Q_2,\ldots,Q_g$ and $\bar P_1,\bar P_2,\ldots,\bar P_g$ on $\gamma$.
\qed

Thus the addition formula (\ref{eq:tepdtl}) on ${\rm Sym}^g(\gamma)$ equivalent to the time evolution (\ref{eq:pdTL}) of the pdTL is realized by using the intersection of $\gamma$, $\kappa_1$, and $\kappa_2$.

\begin{example}
Put $g=3$.
The spectral curve $\gamma$ is given by
\begin{align*}
&f(u,v)=v^2-\left(u^{4}+c_3u^3+c_2u^2+c_1u+c_0\right)^2+4c_{-1},\\
&c_3
=
\sum_{i=1}^4\left(I_i+V_i\right),\quad
c_2
=
\sum_{1\leq i<j\leq4}\left(I_iI_j+V_iV_j\right)+\sum_{i=1}^4I_i\left(V_{i+1}+V_{i+2}\right),\\
&c_1
=
\sum_{1\leq i<j<k\leq4}\left(I_iI_jI_k+V_iV_jV_k\right)+\sum_{i=1}^4I_i\left(I_{i+1}+V_{i+1}\right)V_{i+2},\\
&c_0
=\prod_{i=1}^4I_i+\prod_{i=1}^4V_i,\quad
c_{-1}
=\prod_{i=1}^4I_iV_i.
\end{align*}

The curve $\kappa_1$ passing through $P_1,P_2,P_3$ and $A=(0,\prod_{i=1}^4V_i-\prod_{i=1}^4I_i)$ is given by 
\begin{align*}
h_1(u,v)
= t_{4}(c_0)+ t_{4}(c_1)u+ t_{4}(c_2)u^2+ t_{4}(c_3)u^3+u^4+v.
\end{align*}
The curve $\kappa_2$ passing through $Q_1,Q_2,Q_3$, the intersection points of $\gamma$ and $\kappa_1$, is given by
\begin{align*}
h_2(u,v)
=
c_0
&+
2I_1V_4
\left(
 I_2 I_3 + I_2 V_3 + V_2 V_3 
\right)\\
&
+
\left\{c_1
+ 2 I_1V_4\left( I_2  +  I_3  + V_2 + V_3\right)
\right\}u\\
&
+
\left(c_2+ 2 I_1 V_4\right)u^2
+
c_3u^3+u^4+v.
\end{align*}

Figure \ref{fig:g3curve} shows an example of the intersection $\gamma$, $\kappa_1$, and $\kappa_2$.

\begin{figure}[htbp]
\centering
{
\includegraphics[scale=.8]{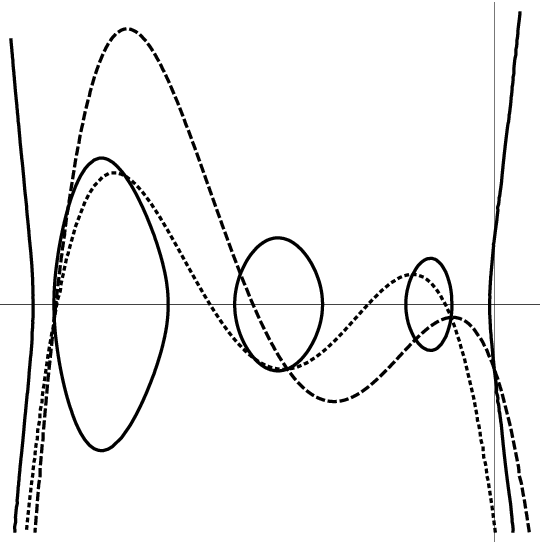}
}
\caption{The solid curve is $\gamma$, the dashed one is $\kappa_1$, and the dotted one is $\kappa_2$.
The intersection point of $\gamma$ and $\kappa_1$ on the vertical axis is $A$.
The triple intersection points of $\gamma$, $\kappa_1$, and $\kappa_2$ are $Q_1$, $Q_2$, and $Q_3$.
The intersection points of $\gamma$ and $\kappa_1$ other than $Q_i$ are $P_1$, $P_2$, and $P_3$.
The intersection points of $\gamma$ and $\kappa_2$ other than $Q_i$ are $\bar P_1$, $\bar P_2$, and $\bar P_3$.
Here we set $I_1=1$, $I_2=3$, $I_3=I_4=2$, $V_1=2$, and $V_2=V_3=V_4=1$.
}
\label{fig:g3curve}
\end{figure}

\end{example}

\subsection{Discrete motion of curves}
We can interpret the time evolution of the pdTL as discrete motion of affine curves.

Let $\mathcal{C}_1$ be the moduli space of the curve $\kappa_1=\left(h_1(u,v)=0\right)$:
\begin{align*}
\mathcal{C}_1
=
\begin{cases}
\left\{\left(s_1(c_0),s_1(c_1),\ldots,s_1(c_g)\right)\right\}&\mbox{for even $g$,}\\
\left\{\left(t_{g+1}(c_0),t_{g+1}(c_1),\ldots,t_{g+1}(c_g)\right)\right\}&\mbox{for odd $g$.}\\
\end{cases}
\end{align*}
Define a map $\tilde\psi: \mathcal{U}_c\to\mathcal{C}_1$ to be
\begin{align*}
\tilde\psi: (I_1,\ldots,I_{g+1},V_1,\ldots,V_{g+1})
\mapsto
\begin{cases}
\left(s_1(c_0),\ldots,s_1(c_g)\right)&\mbox{for even $g$,}\\
\left(t_{g+1}(c_0),\ldots,t_{g+1}(c_g)\right)&\mbox{for odd $g$.}\\
\end{cases}
\end{align*}
Also define a map $\upsilon$ on $\mathcal{C}_1$ to be
\begin{align*}
\upsilon: 
\begin{cases}\left(s_1(c_0),\ldots,s_1(c_g)\right)
\mapsto
\left(s_1(\bar c_0),\ldots,s_1(\bar c_g)\right)&\mbox{for even $g$,}\\
\left(t_{g+1}(c_0),\ldots,t_{g+1}(c_g)\right)
\mapsto
\left(t_{g+1}(\bar c_0),\ldots,t_{g+1}(\bar c_g)\right)&\mbox{for odd $g$}\\
\end{cases}
\end{align*}
so that the following diagram is commutative
\begin{align*}
\begin{CD}
\mathcal{U}_c @> \tilde\psi >> \mathcal{C}_1\\
@V (\ref{eq:pdTL}) VV @VV \upsilon V\\
\mathcal{U}_{c} @>> \tilde\psi > \mathcal{C}_1.\\
\end{CD}
\end{align*}

Then $\upsilon$ induces the discrete motion of curves
\begin{align*}
\kappa_1^0\to \kappa_1^1\to \kappa_1^2\to\cdots,
\end{align*}
where $\kappa_1^t=\left(h_1^t(u,v)=0\right)$ and
\begin{align*}
h_1^t(u,v)
=
\begin{cases}
\DIS
\sum_{i=0}^gs_1(c_i^t)u^i
+
u^{g+1}
-
v
&\mbox{for even $g$},\\
\DIS
\sum_{i=0}^gt_{g+1}(c_i^t)u^i
+
u^{g+1}
+
v&\mbox{for odd $g$}.\\
\end{cases}
\end{align*}

\begin{figure}[htbp]
\centering
{
\includegraphics[scale=.3]{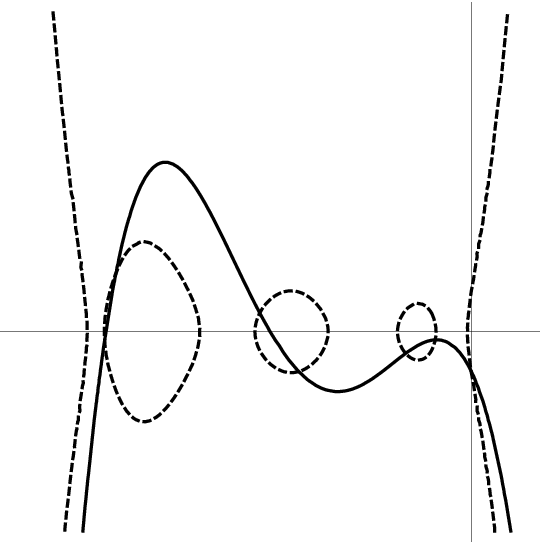}
\includegraphics[scale=.3]{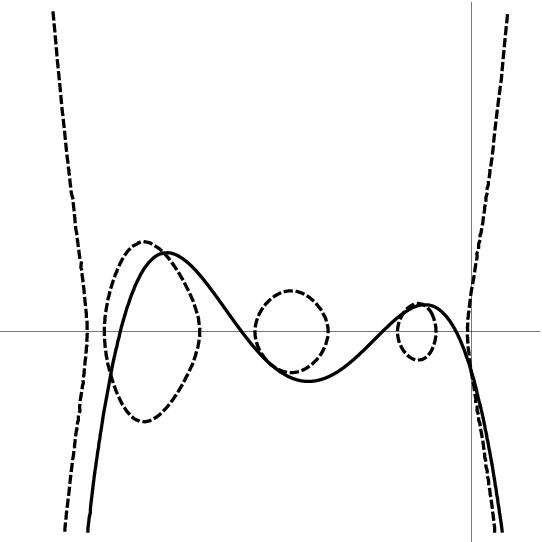}
\includegraphics[scale=.3]{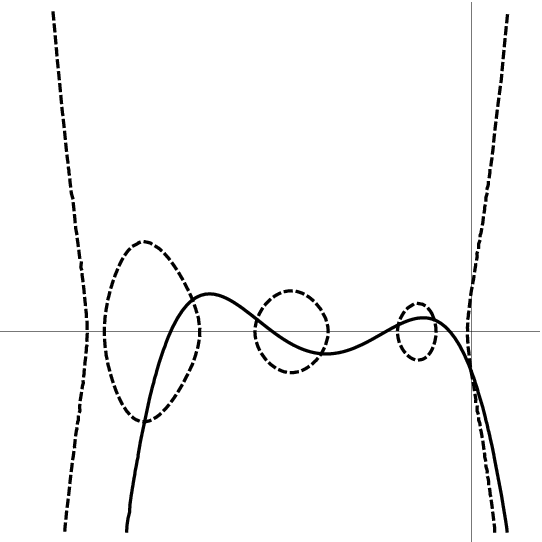}
\includegraphics[scale=.3]{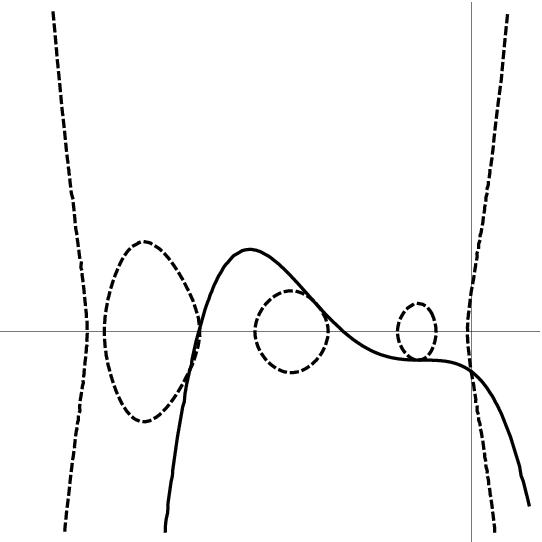}
\\
\includegraphics[scale=.3]{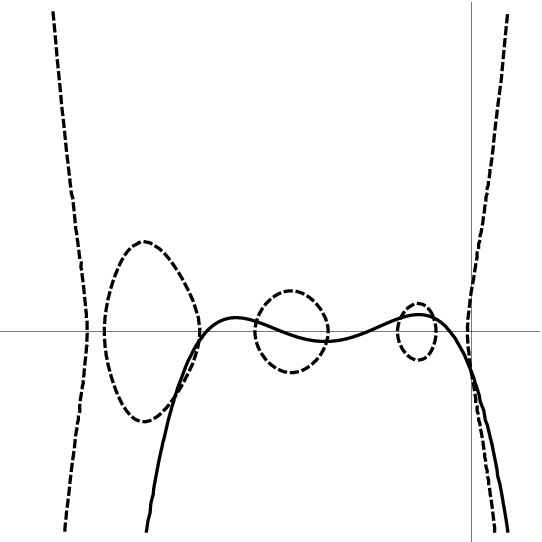}
\includegraphics[scale=.3]{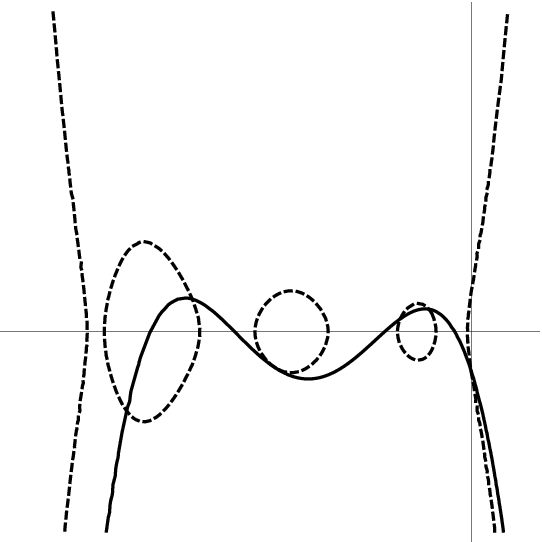}
\includegraphics[scale=.3]{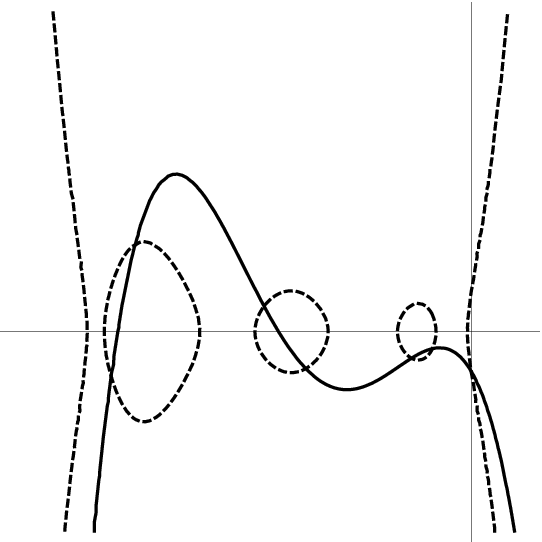}
\includegraphics[scale=.3]{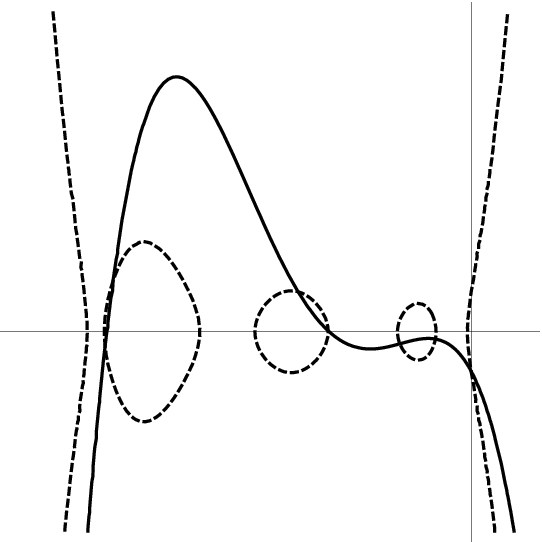}
}
\caption{The discrete motion of the curves $\kappa_1^t$ ($t=0,1,\ldots,7$) induced by $\upsilon$. 
The solid curves are $\kappa_1^t$ and the dashed one is $\gamma$.
The figures are sorted in time increasing order from left to right and top to bottom.
Here we set $I_1^0=1$, $I_2^0=3$, $I_3^0=I_4^0=2$, $V_1^0=2$, and $V_2^0=V_3^0=V_4^0=1$.
}
\label{fig:CMG3}
\end{figure}

Figure \ref{fig:CMG3} shows an example of the discrete motion of $\kappa_1^t$.

\section{Ultradiscrete periodic Toda lattice and tropical hyperelliptic curves}
\label{sec:ATHCPBS}
In this section, we tropicalize the geometric framework concerning the time evolution of the pdTL shown above.
We then present a tropical geometric realization of the ultradiscrete periodic Toda lattice, which is the ultradiscretization of the pdTL.
\subsection{UD-pTL and its spectral curve}
Suppose that the pdTL has a one parameter family of positive solutions $I_i(\epsilon)>0$ and $V_i(\epsilon)>0$ for $i=1,2,\ldots,g+1$, where $\epsilon$ is a positive number.
Also suppose that the limits $\lim_{\epsilon\to+0}-\epsilon\log I_i(\epsilon)=J_i\in\R$ and $\lim_{\epsilon\to+0}-\epsilon\log V_i(\epsilon)=W_i\in\R$ exist.
Then $J_i$ and $W_i$ ($i=1,2,\ldots,g+1$) solve the difference equation
\begin{align}
\bar J_i&=\left\lfloor W_i,X_i+J_i\right\rfloor,\qquad
\overline{W}_i=J_{i+1}+W_i-\bar J_i,\label{eq:pbbs}\\
X_i&=\left\lceil\sum_{l=1}^k\left(J_{i-l}-W_{i-l}\right)\right\rceil_{0\leq k\leq g},\nn
\end{align}
where we define
\begin{align*}
\lfloor A,B,\ldots\rfloor:=\min[A,B,\ldots],
\qquad
\lceil A,B,\ldots\rceil:=\max[A,B,\ldots]
\end{align*}
for $A,B,\ldots\in\R$.
Here we assume that $\sum_{l=1}^0\left(J_{i-l}-W_{i-l}\right)=0$ and 
\begin{align}
\sum_{i=1}^{g+1}J_i<\sum_{i=1}^{g+1}W_i.
\label{eq:jwcond}
\end{align}
We denote the map $\R^{2g+2}\to\R^{2g+2}$;
\begin{align*}
(J_1,\ldots,J_{g+1},W_1,\ldots,W_{g+1})
\mapsto
(\bar J_1,\ldots,\bar J_{g+1},\bar W_1,\ldots,\bar W_{g+1})
\end{align*}
by $\zeta$ and use the notation
\begin{align*}
\left(J_1^t,\ldots,J_{g+1}^t,W_1^t,\ldots,W_{g+1}^t\right)
:=
\underset{t}{\underbrace{\zeta\circ\zeta\circ\cdots\circ\zeta}}(J_1,\cdots,J_{g+1},W_1,\cdots,W_{g+1})
\end{align*}
for $t=0,1,\ldots$.

We call the dynamical system generated by (\ref{eq:pbbs}) the ultradiscrete periodic Toda lattice (UD-pTL).
In particular, if the variables $J_1,\ldots,J_{g+1}$ and $W_1,\ldots,W_{g+1}$ take their values in positive integer then the dynamical system is called the periodic box-ball system (pBBS) \cite{YT02, KT02}.
The pBBS is obtained from the box-ball system (BBS), which was introduced by Takahashi and Satsuma as a soliton cellular automaton \cite{TS90}, by imposing a periodic boundary condition.
The procedure which reduces (\ref{eq:pdTL}) to (\ref{eq:pbbs}) is called the ultradiscretization \cite{TTMS96}.

Now we introduce the spectral curve of the UD-pTL by using the procedure of ultradiscretization.
Note first that all coefficients $c_{-1},c_0,\ldots,c_g$ in $\tilde f$, the defining polynomial of the spectral curve $\tilde\gamma$ of the pdTL, are subtraction-free (see proposition \ref{prop:subfree}). 
Hence we can apply the ultradiscretization procedure to them; suppose that the positive numbers $x$, $y$, and $c_i$ are parametrized with $\epsilon>0$ and the limits $\lim_{\epsilon\to+0}-\epsilon\log x=X$, $\lim_{\epsilon\to+0}-\epsilon\log y=Y$, and $\lim_{\epsilon\to+0}-\epsilon\log c_i=C_i$ exist ($i=-1,0,\ldots,g$). 
Then $\tilde f$ reduces to the following tropical polynomial $F$ in the limit $\epsilon\to+0$
\begin{align*}
F(X,Y)
:=
\left\lfloor
2Y,Y+\left\lfloor(g+1)X,C_g+gX,\cdots,C_1+X,C_0\right\rfloor,C_{-1}
\right\rfloor.
\end{align*}

By definition, the coefficient $C_i$ ($i=-1,0,\ldots,g$) is a tropical polynomial in $J_j$ and $W_k$ ($j,k=1,2,\ldots,g+1$).
Therefore, the correspondence defines a piecewise linear map
\begin{align}
\psi:\R^{2g+2}\to\R^{g+2};\ 
\left(J_1,\ldots,J_{g+1},W_1,\ldots,W_{g+1}\right)
\mapsto
\left(C_{-1},C_0,\ldots,C_g\right).
\label{eq:CJW}
\end{align}
We may use the notation $C_i^t:=C_i\left(J_1^t,\ldots,J_{g+1}^t,W_1^t,\ldots,W_{g+1}^t\right)$ ($t=0,1,\ldots$).

In general, $C_i$ has a complicated form because it is the ultradiscretization of $c_i$ (see (\ref{eq:c1coef})).
For small or large $i$, however, $C_i$ has a relatively simple form:
\begin{align*}
C_g
&=
{\left\lfloor{J_i,W_i}\right\rfloor_{1\leq i\leq g+1}},\\
C_{g-1}
&=
\left\lfloor
\underset{1\leq i< j\leq g+1}{\left\lfloor{J_i+J_j,W_i+W_j}\right\rfloor},
\underset{\substack{1\leq i, j\leq g+1\\j\neq i,i-1}}{\left\lfloor J_i+W_j\right\rfloor}
\right\rfloor,\\
C_0
&=
\left\lfloor
\sum_{i=1}^{g+1}J_i,\sum_{i=1}^{g+1}W_i
\right\rfloor
=
\sum_{i=1}^{g+1}J_i,\\
C_{-1}
&=
\sum_{i=1}^{g+1}\left(J_i+W_i\right).
\end{align*}
Here we use the assumption (\ref{eq:jwcond}).

One can show that the coefficients $C_{-1},C_0,\ldots,C_g$ in the tropical polynomial $F$ are the conserved quantities of the UD-pTL \cite{IT08}.
These conserved quantities can also be constructed via paths on a two dimensional array of boxes \cite{MIT05}.

\begin{figure}[htbp]
\centering
{\unitlength=.05in{\def\arraystretch{1.0}
\begin{picture}(80,64)(-5,16)
\thicklines
\thicklines
\put(51,45){\line(-4,-1){8}}
\put(51,51){\line(-4,1){8}}
\put(51,45){\line(0,1){6}}
\put(43,43){\line(-3,-1){6}}
\put(43,53){\line(-3,1){6}}
\put(43,43){\line(0,1){10}}
\put(37,41){\line(0,1){14}}

\dashline{1}(37,41)(27,36)
\dashline{1}(37,55)(27,60)

\put(27,36){\line(0,1){24}}
\put(27,36){\line(-3,-2){6}}
\put(27,60){\line(-3,2){6}}
\put(21,32){\line(0,1){32}}
\put(21,32){\line(-1,-1){8}}
\put(21,64){\line(-1,1){8}}
\put(13,24){\line(0,1){48}}

\dottedline(66,45)(51,45)
\dottedline(66,51)(51,51)

\dottedline(13,24)(9,16)
\dottedline(13,72)(9,80)
\put(15,22){\makebox(0,0){$V_0^\prime$}}
\put(22,29){\makebox(0,0){$V_1^\prime$}}
\put(51,42){\makebox(0,0){$V_g^\prime$}}
\put(15,73){\makebox(0,0){$V_0$}}
\put(22,66){\makebox(0,0){$V_1$}}
\put(51,54){\makebox(0,0){$V_g$}}
\end{picture}
}}
\caption{
The tropical hyperelliptic curve $\tilde\Gamma$.
The compact subset $\Gamma$ is drown by solid lines and the half rays by dotted lines.
}
\label{fig:THC}
\end{figure}
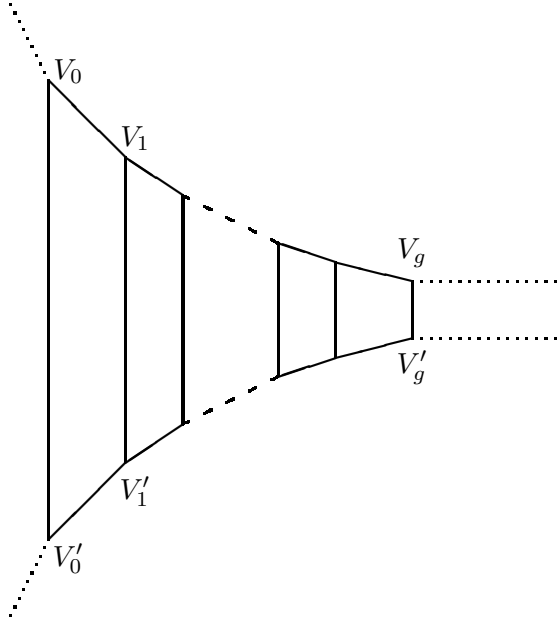

Consider the tropical curve $\tilde\Gamma$ defined by the tropical polynomial $F$ \cite{RGST03,Gathmann06,IMS07}:
\begin{align*}
&\tilde\Gamma
:=
\left\{
P\in\R^2\ |\ 
\mbox{$F$ is not differentiable at $P$}
\right\}.
\end{align*}
Assume $C_{-1}>2C_0$, $C_{g-1}>2C_g$, and $C_i+C_{i+2}>2C_{i+1}$ for $i=0,1,\ldots,g-2$.
Then $\tilde\Gamma$ is a tropical hyperelliptic curve of genus $g$ \cite{MZ06,HMY09}.
The tropical curve $\tilde\Gamma$ is often referred to as the tropicalization of the hyperelliptic curve $\tilde\gamma$ or $\gamma$.
By removing all half rays from $\tilde\Gamma$, we obtain the compact tropical curve denoted by $\Gamma$ (see figure \ref{fig:THC}).
The tropical hyperelliptic curves $\tilde\Gamma$ and $\Gamma$ are called the spectral curves of the UD-pTL.

Consider the involution $\iota:\tilde\Gamma\to\tilde\Gamma; P=(X,Y)\mapsto P^\prime=(X,C_{-1}-Y)$.
We call $P^\prime$ the conjugate of $P$.
Note that $\tilde\Gamma$ is symmetric with respect to the line $Y={C_{-1}}/{2}$.

We denote the $2g+2$ vertices of $\Gamma$ by $V_i$ and $V_i^\prime$ for $i=0,1,\ldots,g$
\begin{align*}
&V_i
=
\left(
C_{g-i}-C_{g-i+1}, C_{-1}-(g-i+1)C_{g-i}+(g-i)C_{g-i+1}
\right),\\
&V_i^\prime=
\left(
C_{g-i}-C_{g-i+1}, (g-i+1)C_{g-i}-(g-i)C_{g-i+1}
\right).
\end{align*}
The cycle connecting $V_i$, $V_{i-1}$, $V_{i-1}^\prime$, and $V_i^\prime$ in a counterclockwise direction is denoted by $\alpha_i$ for $i=1,2,\ldots,g$.

\subsection{Addition on tropical hyperelliptic curves}
We briefly review addition of points on tropical hyperelliptic curves \cite{Vigeland04,Nobe12}.

Denote the divisor group of the tropical hyperelliptic curve $\tilde\Gamma$ by $\mathcal{D}(\tilde\Gamma)$ as in the non-tropical case (To reduce symbols, we often use the same ones as in the non-tropical case).
A rational function on $\tilde\Gamma$ is a continuous function $f:\tilde\Gamma\to\R\cup\{\pm\infty\}$ such that its restriction to any edge is piecewise linear with integral slope \cite{Gathmann06}.
The order of $f$ at $P\in\tilde\Gamma$ is the sum of the outgoing slopes of all segments emanating from $P$ and is denoted by ${\rm ord}_Pf$.
If ${\rm ord}_Pf<0$ then $P$ is called the zero of $f$ of order $\left|{\rm ord}_Pf\right|$.
If ${\rm ord}_Pf>0$ then $P$ is called the pole of $f$ of order ${\rm ord}_Pf$.
This is because we choose the tropical semi-field $\mathbb{T}=\mathbb{R}\cup\{\infty\}$ equipped with the operations $\min$ and $+$.
The principal divisor $(f)$ of $f$ is defined to be $(f):=\sum_{P\in\tilde\Gamma}\left({\rm ord}_Pf\right)P$.
We then find $\deg (f)=0$.

Define the Picard group of $\tilde\Gamma$ to be the residue class group ${\rm Pic}^0(\tilde\Gamma):=\mathcal{D}_0(\tilde\Gamma)\slash\mathcal{D}_l(\tilde\Gamma)$, where $\mathcal{D}_0(\tilde\Gamma)$ is the group of divisors of degree 0 on $\tilde\Gamma$ and $\mathcal{D}_l(\tilde\Gamma)$ is the group of principal divisors of rational functions on $\tilde\Gamma$.
We also define ${\rm Pic}^0(\Gamma):=\mathcal{D}_0(\Gamma)\slash\mathcal{D}_l(\Gamma)$ for the compact subset $\Gamma$ of $\tilde\Gamma$.
We then find the following \cite{GK06,Vigeland04,Nobe12}
\begin{align*}
{\rm Pic}^0(\tilde\Gamma)
=
{\rm Pic}^0(\Gamma).
\end{align*}

Let us define $L(D):=\left\{k\in\R(\Gamma)\ |\ (k)+D>0\right\}$ for $D\in\mathcal{D}(\Gamma)$, where $\R(\Gamma)$ is the field of rational functions on $\Gamma$ \cite{GK06,BN07}.
Also define the rank of $L(D)$ to be the maximal integer $k$ such that for all choices of (not necessarily distinct) points $P_1,P_2,\ldots,P_k\in\Gamma$ we have $L(D-P_1-P_2-\cdots-P_k)\neq\emptyset$ \cite{GK06}.
We denote it by ${\rm rank}\mkern2mu L(D)$.
The following theorem (a corollary of the tropical Riemann-Roch theorem) is useful \cite{GK06,MZ06,BN07}.
\begin{theorem}\label{crl:CorRRT}
For any divisor $D$ on a tropical hyperelliptic curve $\Gamma$ of genus $g$ such that $\deg D>2g-2$, we have
\begin{align*}
{\rm rank}\mkern2mu L(D)=\deg D-g.
\end{align*}
\qed
\end{theorem}

As in the non-tropical case, we define the canonical map $\Phi:\mathcal{D}_g^+(\Gamma)\to {\rm Pic}^0(\Gamma)$ to be
\begin{align*}
\Phi(A)
:\equiv
A-D^\ast
\quad
\mbox{(mod $\mathcal{D}_l(\Gamma)$)}
\qquad
\mbox{for $A\in\mathcal{D}_g^+(\Gamma)$},
\end{align*}
where $\mathcal{D}_g^+(\Gamma)$ is the group of effective divisors of degree $g$ on $\Gamma$ and $D^\ast\in\mathcal{D}_g^+(\Gamma)$ is a fixed element.
We then have the following theorem.
\begin{theorem}[\cite{Nobe12}]
The canonical map $\Phi$ is surjective.
In particular, $\Phi$ is bijective if $g=1$.
\qed
\end{theorem}

By using the surjection $\Phi$, we induce addition of points on the $g$-th symmetric product ${\rm Sym}^g(\Gamma):=\Gamma^g/\mathfrak{S}_g$ from ${\rm Pic}^0(\Gamma)$.
Put $\tilde\Phi:=\Phi\circ\mu^{-1}:{\rm Sym}^g(\Gamma)\to{\rm Pic}^0(\Gamma)$, where $\mu:\mathcal{D}_g^+(\Gamma)\to{\rm Sym}^g(\Gamma); D_P=P_1+P_2+\cdots+P_g\mapsto d_P:=\mu(D_P)=\left\{P_1,P_2,\ldots,P_g\right\}$.
For $d_P, d_Q\in{\rm Sym}^g(\Gamma)$, we define $d_P\oplus d_Q$ to be an element in the subset
\begin{eqnarray*}
\tilde\Phi^{-1}\left(\tilde\Phi(d_P)+\tilde\Phi(d_Q)\right)
\subset{\rm Sym}^g(\Gamma).
\end{eqnarray*}

Put $\alpha_{ij}:=\alpha_i\cap\alpha_j\setminus\left\{\mbox{end points of $\alpha_i\cap\alpha_j$}\right\}$ for the cycles $\alpha_i$ ($i=1,2,\ldots,g$).
We define the subset $\tilde{\mathcal{D}}$ of $\mathcal{D}_{g}^{+}(\Gamma)$ to be
\begin{align*}
\tilde{\mathcal{D}}
:=
\left\{
\begin{array}{l}
D_P\in\mathcal{D}_{g}^{+}(\Gamma)\\
\end{array}
\
\left|
\begin{array}{l}
\mbox{$P_i\in\alpha_i$ for all $i=1,2,\ldots,g$ and} \\
\mbox{there exists at most one point on $\alpha_{ij}$}
\end{array}
\right.
\right\}.
\end{align*}
We then have the following theorem.
\begin{theorem}[\cite{IT08}]\label{thm:bijection}
The reduced map $\Phi|{\tilde{\mathcal{D}}}:\tilde{\mathcal{D}}\overset{\sim}{\to}{\rm Pic}^0(\Gamma)$ is bijective.
\qed
\end{theorem}

Hereafter, we fix $D^\ast$ as follows
\begin{align*}
D^\ast
=
\begin{cases}
\displaystyle\frac{g}{2}(V_0+V_0^\prime)&\mbox{for even $g$,}\\
\displaystyle\frac{g-1}{2}(V_0+V_0^\prime)+V_0&\mbox{for odd $g$.}\\
\end{cases}
\end{align*}
Define the element $o\in{\rm Sym}^g(\Gamma)$ to be
\begin{align*}
o
:=
\begin{cases}
\DIS\bigcup_{i=1}^{g/2}\left\{V_{2i-1},V_{2i-1}^\prime\right\}&\mbox{for even $g$},\\
\{V_0\}\cup\DIS\left(\bigcup_{i=1}^{g-1/2}\left\{V_{2i},V_{2i}^\prime\right\}\right)&\mbox{for odd $g$.}\\
\end{cases}
\end{align*}
Also define
\begin{align*}
\mathcal{O}
:=
\mu^{-1}(o)
=
\begin{cases}
\DIS\sum_{i=1}^{g/2}\left(V_{2i-1}+V_{2i-1}^\prime\right)&\mbox{for even $g$},\\
V_0+\DIS\sum_{i=1}^{g-1/2}\left(V_{2i}+V_{2i}^\prime\right)&\mbox{for odd $g$.}\\
\end{cases}
\end{align*}
One can show that $\mathcal{O}\in\tilde{\mathcal{D}}\cap\ker\Phi$ holds \cite{Nobe12}.
Therefore, the element $o$ is the unit of addition of the group $\mu(\tilde{\mathcal{D}})\simeq{\rm Pic}^0(\Gamma)$.

Let $d_P$, $d_Q$, and $d_R$ be the elements of $\mu(\tilde{\mathcal{D}})$ satisfying the addition formula
\begin{align*}
d_P
\oplus
d_Q
\oplus
d_R
=
o.
\end{align*}
This can be written by the divisors on ${\rm Pic}^0(\Gamma)$:
\begin{equation*}
D_P+D_Q+D_R-3D^\ast
\equiv0
\quad
\mbox{(mod $\mathcal{D}_l(\Gamma)$)}.
\end{equation*}
There exists a rational function $k\in L(3D^\ast)$ whose $3g$ zeros are $P_1,\ldots,P_g$, $Q_1,\ldots,Q_g$, $R_1,\ldots,R_g$.
If the rational function $k$ is given by a tropical polynomial then we can define a tropical curve $C$ as the set of points at which the tropical polynomial is not differentiable.
Then the above addition is realized by using the intersection of $\Gamma$ and $C$ \cite{Nobe12}.

\subsection{Tropical Jacobians}
Tropical Jacobians were introduced by Mikhalkin and Zharkov in 2006 \cite{MZ06}.
Let $\mathcal{E}(\Gamma)$ be the set of edges of $\Gamma$.
Define the weight $w:\mathcal{E}(\Gamma)\to\R_{\geq0}$ by
\begin{align*}
w(e)
=
\frac{\| e\|}{\| \xi_e\|},
\end{align*}
where $\xi_e$ is the primitive tangent vector of $e\in\mathcal{E}(\Gamma)$ and $\|\ \|$ denotes the Euclidean norm in $\R^2$.
We define a symmetric bilinear form $Q$ on the space of paths in $\Gamma$ as follows.
For a non-self-intersecting path $\varpi$, set $Q(\varpi,\varpi):=(\mbox{length of $\varpi$ with respect to $w$})$, and extend it to any pairs of paths bilinearly.

\begin{definition}
The tropical Jacobian of $\Gamma$ is a $g$-dimensional real torus defined to be
\begin{align*}
J(\Gamma)
:=
\R^g\slash \bLambda\Z^g,
\end{align*}
where $\bLambda=(\Lambda_{ij})$ is the $g\times g$ real matrix given by
\begin{align*}
\Lambda_{ij}
=
Q\left(
\sum_{k=1}^i\alpha_k,
\sum_{l=1}^j\alpha_l
\right)
=
C_{-1}
+
r_i\delta_{ij}
+
2\lfloor\lambda_i,\lambda_j\rfloor,\\
\lambda_i
=
C_{g-i}
-
C_{g-i+1},\qquad
r_i
=
C_{-1}
-
2\sum_{k=1}^g\lfloor\lambda_i,\lambda_k\rfloor
\end{align*}
for $i,j=1,2,\ldots,g$.
\end{definition}

Fix a point $P_0$ on $\Gamma$.
Let the path from $P_0$ to $P_i$ on $\Gamma$ be $\varpi_i$.
We define the Abel-Jacobi map $\eta:\mathcal{D}_g^+(\Gamma)\to J(\Gamma)$ to be
\begin{align*}
\eta:
D_P
=
P_1+\cdots+P_g
\mapsto
\sum_{i=1}^g\left(Q(\varpi_i,\alpha_1),\cdots, Q(\varpi_i,\alpha_g)\right).
\end{align*}

\subsection{Time evolution}
Let  the phase space of the UD-pTL be $\mathcal{T}:=\left\{(J_1,\ldots,J_{g+1},W_1,\ldots,W_{g+1})\ |\ \sum_{i=1}^{g+1}J_i<\sum_{i=1}^{g+1}W_i\right\}$ and the moduli space of $\Gamma$ be $\mathcal{C}:=\left\{(C_{-1},C_0,\ldots,C_g)\right\}$.
Consider the map $\psi:\mathcal{T}\to\mathcal{C}$ defined by (\ref{eq:CJW}) and set 
\begin{align*}
\mathcal{T}_C
:=
\psi^{-1}(C_{-1},C_0,\ldots,C_g)\subset\mathcal{T}.
\end{align*}
Let $\phi:\mathcal{T}\to \mathcal{D}_g^+(\Gamma)$ be the tropical eigenvector map \cite{IT08}.
Then the UD-pTL is linearized on the tropical Jacobian $J(\Gamma)$.
\begin{theorem}[\cite{IT08,IT09}]
Define the transformation operator $\nu$ to be
\begin{align*}
\nu:
J(\Gamma)\to J(\Gamma);\
\bold{z}
\mapsto
\bold{z}
+
\left(\lambda_1,\lambda_2-\lambda_1,\ldots,\lambda_g-\lambda_{g-1}\right).
\end{align*}
Then the following diagram is commutative
\begin{align*}
\begin{CD}
\mathcal{T}_C @> \eta\circ\phi >>J(\Gamma)\\
@V (\ref{eq:pbbs}) VV @VV \nu V\\
\mathcal{T}_C @>> \eta\circ\phi > J(\Gamma).\\
\end{CD}
\end{align*}
\qed
\end{theorem}

Define $T\in\mathcal{D}_g^+(\Gamma)$ to be
\begin{align}
T
=
\begin{cases}
\DIS
V_{0}
+
V_g^\prime
+
\sum_{i=1}^{(g-2)/2}\left(V_{2i}+V_{2i}^\prime\right)
&
\mbox{for even $g$,}\\
\DIS
B
+
V_1
+
V_g^\prime
+
\sum_{i=2}^{(g-1)/2}\left(V_{2i-1}+V_{2i-1}^\prime\right)
&
\mbox{for odd $g$}\\
\end{cases}
\label{eq:tepBBS}
\end{align}
for $g\geq2$.
Here $B=(C_g,C_{-1}-C_{g-1}-(g-1)C_g)\in\overline{V_0V_0^\prime}\subset\alpha_1$ is the unique point such that $B+V_1-2V_0$ is the principal divisor of a rational function on $\Gamma$. 
Note that $Q(\overline{V_0B},\overline{V_0B})=\lambda_1$ and $T\in\tilde{\mathcal D}$.
For an even number $g$, we find
\begin{align*}
\Phi\left(T\right)
&=
V_g^\prime-V_0^\prime
+
\sum_{i=0}^{(g-2)/2}\left(V_{2i}+V_{2i}^\prime\right)
-
\frac{g}{2}\left(V_0+V_0^\prime\right)\\
&\equiv
V_g^\prime-V_0^\prime
\quad
\mbox{(mod $\mathcal{D}_l(\Gamma)$).}
\end{align*}
For an odd number $g$, we also find
\begin{align*}
\Phi\left(T\right)
&=
V_g^\prime-V_0^\prime
+
B+V_1-2V_0
+
\sum_{i=2}^{(g-1)/2}\left(V_{2i-1}+V_{2i-1}^\prime\right)
-
\frac{g-3}{2}\left(V_0+V_0^\prime\right)\\
&\equiv
V_g^\prime-V_0^\prime
\quad
\mbox{(mod $\mathcal{D}_l(\Gamma)$).}
\end{align*}

For $g=1$, we define $T$ to be the point $(C_1,C_0)\in\tilde{\mathcal{D}}$ on the edge $\overline{V_0V_0^\prime}$.
Since there exists a rational function whose principal divisor is $V_0+V_1^\prime-V_0^\prime-T$, we find
\begin{align*}
\Phi(T)
=
T-V_0
\equiv
V_1^\prime-V_0^\prime
\quad
\mbox{(mod $\mathcal{D}_l(\Gamma)$).}
\end{align*}
Thus we obtain the following proposition.
\begin{proposition}\label{prop:phiT}
For any $g$, we have
\begin{align*}
\Phi\left(T\right)
\equiv
V_g^\prime-V_0^\prime
\quad
(\mbox{mod $\mathcal{D}_l(\Gamma)$}).
\end{align*}\qed
\end{proposition}

We obtain the following theorem concerning the time evolution of the UD-pTL and an addition on ${\rm Sym}^g(\Gamma)$.
\begin{theorem}\label{thm:tepBBSonSymg}
Set $\tau=\mu(T)\in {\rm Sym}^g(\Gamma)$.
Then, for any $g$, the following diagram is commutative
\begin{align*}
\begin{CD}
\mathcal{T}_C @> \mu\circ\phi >>{\rm Sym}^g(\Gamma)\\
@V (\ref{eq:pbbs}) VV @VV \oplus\tau V\\
\mathcal{T}_C @>> \mu\circ\phi > {\rm Sym}^g(\Gamma).\\
\end{CD}
\end{align*}
\end{theorem}

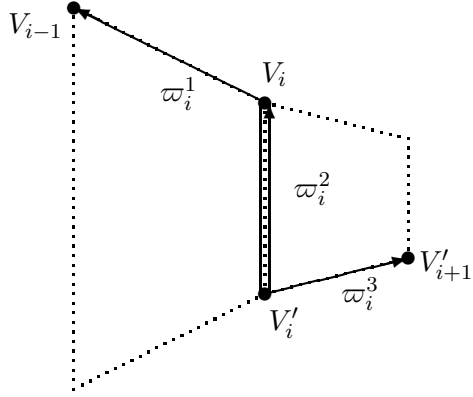
\begin{figure}[htbp]
\centering
{\unitlength=.05in{\def\arraystretch{1.0}
\begin{picture}(80,40)(-23,1)
\thicklines
\dottedline(35,26.25)(35,13.75)
\dottedline(20,30)(35,26.25)
\dottedline(20,10)(35,13.75)
\dottedline(0,40)(20,30)
\dottedline(20,10)(20,30)
\dottedline(0,0)(20,10)
\dottedline(0,0)(0,40)
\put(20.5,10){\vector(0,1){20}}
\put(20,10){\vector(4,1){15}}
\put(19.5,9.75){\line(0,1){20.5}}
\put(19.5,30.25){\vector(-2,1){19.5}}
\put(-4,38){\makebox(0,0){$V_{i-1}$}}
\put(22,7){\makebox(0,0){$V_{i}^\prime$}}
\put(21,33){\makebox(0,0){$V_{i}$}}
\put(39,13){\makebox(0,0){$V_{i+1}^\prime$}}
\put(11,31){\makebox(0,0){$\varpi_{i}^1$}}
\put(25,21){\makebox(0,0){$\varpi_{i}^2$}}
\put(30,10){\makebox(0,0){$\varpi_{i}^3$}}
\put(20,10){\circle*{1.5}}
\put(20,30){\circle*{1.5}}
\put(0,40){\circle*{1.5}}
\put(35,13.75){\circle*{1.5}}
\end{picture}
}}
\caption{
The paths $\varpi_{i}^1$, $\varpi_{i}^2$, and $\varpi_{i}^3$.
}
\label{fig:paths}
\end{figure}

(Proof)\quad
Let the paths from $V_{i}^\prime$ to $V_{i-1}$, $V_{i}$, and $V_{i+1}^\prime$ be $\varpi_{i}^1$, $\varpi_{i}^2$, and $\varpi_{i}^3$ as in  figure \ref{fig:paths}, respectively ($i=1,2,\ldots,g-1$).
Also let the paths from the fixed point $P_0$ to $V_{i-1}$, $V_{i}$, and $V_{i+1}^\prime$ via $\varpi_{i}^1$, $\varpi_{i}^2$, and $\varpi_{i}^3$ be $\widetilde{\varpi}_{i}^1$, $\widetilde{\varpi}_{i}^2$, and $\widetilde{\varpi}_{i}^3$, respectively.

Let $\Pi_S$ be the set of paths on $\Gamma$ emanating from $S\in\Gamma$.
Define $\iota_S:\Pi_S\to\R^g; \omega\mapsto \iota_S(\omega)=\left(Q(\omega,\alpha_i)\right)_{1\leq i\leq g}$.
For $D_P=P_1+\cdots+P_g\in \tilde{\mathcal{D}}$, we see $\eta(D_P)\equiv\sum_{i=1}^g\iota_{P_0}(\omega_i)$ (mod $\bLambda\Z^g$), where $\omega_i$ is a path from $P_0$ to $P_i$.
Let us denote $\iota_{P_0}(\overrightarrow{P_0V_{i}^\prime})$ by $\bold{p}_{i}$, where $\overrightarrow{P_0V_{i}^\prime}$ is a path from $P_0$ to $V_i^\prime$.
Also let the $i$-th standard vector in $\R^g$ be $\bold{e}_i$.
Then we have
\begin{align}
&\iota_{P_0}(\widetilde{\varpi}_{i}^1)
=
\left(r_{i}+\lambda_{i}-\lambda_{i-1}\right)\bold{e}_{i}
-
r_{i}\bold{e}_{i+1}
+
\bold{p}_{i},\nn\\
&\iota_{P_0}(\widetilde{\varpi}_{i}^2)
=
r_{i}\bold{e}_{i}
-
r_{i}\bold{e}_{i+1}
+
\bold{p}_{i},\label{eq:vpi1123}\\
&\iota_{P_0}(\widetilde{\varpi}_{i}^3)
=
\left(\lambda_{i+1}-\lambda_{i}\right)\bold{e}_{i+1}
+
\bold{p}_{i}.\nn
\end{align}

Now assume that $g$ is an even number. 
Then we have
\begin{align*}
&\mathcal{O}
=
\sum_{i=1}^{g/2}\left(V_{2i-1}+V_{2i-1}^\prime\right),\quad
T
=
V_{0}
+
V_g^\prime
+
\sum_{i=1}^{(g-2)/2}\left(V_{2i}+V_{2i}^\prime\right).
\end{align*}
By (\ref{eq:vpi1123}),  we find
\begin{align*}
\eta(\mathcal{O})
=
\sum_{i=1}^{g/2}\left(\iota_{P_0}(\widetilde\varpi_{2i-1}^2)+\bold{p}_{2i-1}\right)
=
\sum_{i=1}^{g/2}r_{2i-1}\left(\bold{e}_{2i-1}-\bold{e}_{2i}\right)+2\sum_{i=1}^{g/2}\bold{p}_{2i-1}
\end{align*}
and
\begin{align*}
&\eta(T)
=
\sum_{i=1}^{g/2}\left(\iota_{P_0}(\widetilde\varpi_{2i-1}^1)+\iota_{P_0}(\widetilde\varpi_{2i-1}^3)\right)\\
&\quad=
\sum_{i=1}^{g/2}\left\{\left(r_{2i-1}+\lambda_{2i-1}-\lambda_{2i-2}\right)\bold{e}_{2i-1}
-
\left(r_{2i-1}-\lambda_{2i}+\lambda_{2i-1}\right)\bold{e}_{2i}\right\}
+
2\sum_{i=1}^{g/2}\bold{p}_{2i-1}.
\end{align*}
Thus we obtain
\begin{align*}
\eta(T)
-
\eta(\mathcal{O})
&=
\sum_{i=1}^{g/2}\left\{\left(\lambda_{2i-1}-\lambda_{2i-2}\right)\bold{e}_{2i-1}
+
\left(\lambda_{2i}-\lambda_{2i-1}\right)\bold{e}_{2i}\right\}\\
&=
\sum_{i=1}^{g}\left(\lambda_{i}-\lambda_{i-1}\right)\bold{e}_{i}.
\end{align*}

Next assume that $g$ is an odd number.
Then we have
\begin{align*}
&\mathcal{O}
=
V_0
+
\sum_{i=1}^{(g-1)/2}\left(V_{2i}+V_{2i}^\prime\right),\quad
T
=
B
+
V_{1}
+
V_g^\prime
+
\sum_{i=2}^{(g-1)/2}\left(V_{2i-1}+V_{2i-1}^\prime\right).
\end{align*}
We find
\begin{align*}
\eta(\mathcal{O})
&=
\sum_{i=1}^{(g-1)/2}\left(\iota_{P_0}(\widetilde\varpi_{2i}^2)+\bold{p}_{2i}\right)
+
\eta(\widetilde\varpi_{0}^2)\\
&=
\sum_{i=1}^{(g-1)/2}r_{2i}\left(\bold{e}_{2i}-\bold{e}_{2i+1}\right)
-
C_{-1}\bold{e}_1
+
2\sum_{i=1}^{(g-1)/2}\bold{p}_{2i}
+
\bold{p}_0
\end{align*}
and
\begin{align*}
\eta(T)
&=
\sum_{i=1}^{(g-1)/2}\left(\iota_{P_0}(\widetilde\varpi_{2i}^1)+\iota_{P_0}(\widetilde\varpi_{2i}^3)\right)
+
\iota_{P_0}(\overrightarrow{P_0B})
+
\bold{p}_0\\
&=
\sum_{i=1}^{(g-1)/2}
\left\{\left(r_{2i}+\lambda_{2i}-\lambda_{2i-1}\right)\bold{e}_{2i}
-
\left(r_{2i}-\lambda_{2i+1}+\lambda_{2i}\right)\bold{e}_{2i+1}\right\}\\
&\qquad-
\left(C_{-1}-\lambda_1\right)\bold{e}_1
+
2\sum_{i=1}^{(g-1)/2}\bold{p}_{2i}+\bold{p}_0.
\end{align*}
It follows that we have
\begin{align*}
\eta(T)
-
\eta(\mathcal{O})
&=
\sum_{i=1}^{(g-1)/2}\left\{\left(\lambda_{2i}-\lambda_{2i-1}\right)\bold{e}_{2i}
+
\left(\lambda_{2i+1}-\lambda_{2i}\right)\bold{e}_{2i+1}\right\}
+
\lambda_1\bold{e}_1\\
&=
\sum_{i=1}^{g}\left(\lambda_{i}-\lambda_{i-1}\right)\bold{e}_{i}.
\end{align*}
Since $\mu:\mathcal{D}_g^+(\Gamma)\to{\rm Sym}^g(\Gamma)$ and $\eta|\tilde{\mathcal{D}}: \tilde{\mathcal{D}}\to J(\Gamma)$ are bijective, this completes the proof. \qed

\section{A geometric realization of UD-pTL}
\label{sec:AGROP}
\subsection{Ultradiscretization of rational functions}
In order to tropicalize the geometric framework of the pdTL constructed in section \ref{sec:grpdtl}, we first introduce a rational map which maps the defining polynomials of the curves appearing in the geometric realization into subtraction-free ones.
Then we apply the ultradiscretization procedure to them.

Let $\sigma:\C_{(u,v)}^2\to\C_{(x,y)}^2$ be the rational map
\begin{align*}
(x,y)
=
\sigma(u,v)
=
\left(
u,\frac{v-u^{g+1}-c_gu^g-\cdots-c_1u-c_0}{2}
\right).
\end{align*}
If we apply $\sigma$ to the points on $\gamma$ then we recover $\tilde\gamma$. 
Therefore, $\sigma$ is the inverse of $\rho$ on $\gamma$.
Note that the defining function $\tilde f$ of $\tilde\gamma$ is subtraction-free.

Consider the curve $\kappa_1=\left(h_1(u,v)=0\right)$ introduced in \S\ref{sec:acptgp}.
Applying $\sigma$ to the points on $\kappa_1$, we obtain an affine curve 
\begin{align*}
{\tilde\kappa_1}
:=
\left(
\tilde h_1(x,y)
=
0
\right)
=
\left\{
(x,y)
=
\sigma(u,v)\ \left|\
h_1(u,v)=0
\right.
\right\},
\end{align*}
where 
\begin{align*}
&\tilde h_1(x,y)
=
\begin{cases}
\DIS
\sum_{i=0}^g( s_1(c_i)-c_i)x^i
-
2y&\mbox{for even $g$,}\\
\DIS
\sum_{i=0}^g( t_{g+1}(c_i)+c_i)x^i
 +
2x^{g+1}
+
2y&\mbox{for odd $g$.}
\end{cases}
\end{align*}
We see that $\tilde h_1$ for an odd number $g$ is subtraction-free as well as $\tilde f$.
Similarly, for an even number $g$, $-\tilde h_1$ is subtraction-free.
Therefore, we can simultaneously apply the procedure of ultradiscretization (or tropicalization with trivial valuation \cite{Maclagan12}) to the rational functions $\tilde f$ and $\tilde h_1$.
Thus we recover the intersection of $\gamma$ and $\kappa_1$ in terms of the tropical curves $\Gamma$ and $K_1$, which is defined below, in the framework of tropical geometry.

Now we introduce the transformations $S_j$ and $T_k$ on the moduli space $\mathcal{C}=\{(C_{-1},C_0,\ldots,C_g)\}$ of $\Gamma$:
\begin{align*}
&S_j(C_i)
:=
\lim_{\epsilon\to0}-\epsilon\log(c_i- s_j(c_i))
=
C_i(\infty,\ldots,\infty,J_j,\infty,\ldots,\infty),\\
&T_k(C_i)
:=
\lim_{\epsilon\to0}-\epsilon\log( t_k(c_i)+c_i)
=
C_i(J_1,\ldots,W_{k-1},\infty,W_{k+1},\ldots,W_{g+1})
\end{align*}
for $i=-1,0,\ldots,g$ and $j,k=1,2,\ldots,g+1$.
Note that $S_j$ eliminates the terms in $C_i$ not containing $J_j$ and $T_k$ the terms in $C_i$ containing $W_k$.
By applying the ultradiscretization procedure to $\tilde h_1$, we obtain the tropical polynomial
\begin{align}
&H_1(X,Y)
=
\begin{cases}
\left\lfloor
\underset{0\leq i\leq g}{\left\lfloor S_1(C_i)+iX\right\rfloor},Y
\right\rfloor
&\mbox{for even $g$,}\\
\left\lfloor
\underset{0\leq i\leq g}{\left\lfloor T_{g+1}(C_i)+iX\right\rfloor},(g+1)X,Y
\right\rfloor
&\mbox{for odd $g$.}
\end{cases}
\label{eq:trophk}
\end{align}
In order to consider non-degenerate curves, we assume 
\begin{align*}
\begin{cases}
S_1(C_i)+S_1(C_{i+2})>2S_1(C_{i+1})
&\mbox{for even $g$,}\\
T_{g+1}(C_i)+T_{g+1}(C_{i+2})>2T_{g+1}(C_{i+1})
&\mbox{for odd $g$}\\
\end{cases}
\end{align*}
for $i=0,1,\ldots,g-2$.

We define the tropical curve $K_1$ to be the set of points at which $H_1$ is not differentiable:
\begin{align*}
K_1
:=
\left\{
P\in\R^2\ |\ 
\mbox{$H_1$ is not differentiable at $P$}
\right\}.
\end{align*}

\subsection{Tropical curve intersections}
We show that the intersection of $\Gamma$ and $K_1$ realizes an addition on ${\rm Sym}^g(\Gamma)$.
\begin{lemma}\label{lem:troph}
The restriction $H_1|\Gamma$ of $H_1$ on $\Gamma$ satisfies $H_1|\Gamma\in L(V_0^\prime+2D^\ast)$ for any $g$.
\end{lemma}

(Proof)\quad
Note first that we have
\begin{align*}
L(V_0^\prime+2D^\ast)
=
\begin{cases}
L\left(gV_0+(g+1)V_0^\prime\right)&\mbox{for even $g$,}\\
L\left((g+1)V_0+gV_0^\prime\right)&\mbox{for odd $g$.}
\end{cases}
\end{align*}
By the tropical Riemann-Roch theorem (theorem \ref{crl:CorRRT}), we find
\begin{align*}
{\rm rank}\mkern2mu L(V_0^\prime+2D^\ast)
=
g+1.
\end{align*}
Also note that the principal divisors of the coordinate functions $x$ and $y$ are given by
\begin{align*}
(x)
=
V_g+V_g^\prime-V_0-V_0^\prime,\qquad
(y)
=
(g+1)V_0-(g+1)V_0^\prime.
\end{align*}

If $g$ is an even number then the $g+2$ rational functions $1,x,,\ldots,gx,y$ on $\Gamma$ are in $L(V_0^\prime+2D^\ast)$.
Similarly, if $g$ is an odd number then the $g+2$ rational functions $1,x,\ldots,gx$, $\left\lfloor (g+1)x, y\right\rfloor$ on $\Gamma$ are in $L(V_0^\prime+2D^\ast)$.
Here we use the fact that the rational function $\left\lfloor (g+1)x, y\right\rfloor$ has a pole of order $g$ at $V_0^\prime$ and that of order $g+1$ at $V_0$.
We denote the tropical module spanned by these rational functions by $\mathcal{M}_1$ \cite{MZ06}:
\begin{align*}
\mathcal{M}_1
:=
\begin{cases}
\left\langle
1,x,\ldots,gx,y
\right\rangle
&\mbox{for even $g$,}\\
\left\langle
1,x,\ldots,gx,\left\lfloor (g+1)x, y\right\rfloor
\right\rangle
&\mbox{for odd $g$.}
\end{cases}
\end{align*}
Then $\mathcal{M}_1\subset L(V_0^\prime+2D^\ast)$. 
Since the rank of $\mathcal{M}_1$ is $g+2$ as a tropical module and ${\rm rank}\mkern2mu L(V_0^\prime+2D^\ast)=g+1$, $\mathcal{M}_1$ is the maximal tropical module in $L(V_0^\prime+2D^\ast)$.
Clearly, $H_1|\Gamma\in \mathcal{M}_1$ holds for any $g$.
\qed

\begin{lemma}\label{lem:troph2}
The tropical curve $K_1$ passes through the vertex $V_g^\prime=(C_0-C_1,C_0)$ of $\Gamma$.
\end{lemma}

(Proof)\quad
Note that we have
\begin{align*}
c_0- s_1(c_0)
=
2\prod_{i=1}^{g+1}I_i,\qquad
 t_{g+1}(c_0)+c_0
=
2\prod_{i=1}^{g+1}I_i.
\end{align*}
This implies that we have
\begin{align*}
 S_1(C_0)
=
 T_{g+1}(C_0)
=
C_0
=
\sum_{i=1}^{g+1}J_i.
\end{align*}

Now assume that $g$ is an even number.
There exists an infinite edge $E$ of $K_1$ defined by $ S_1(C_0)=Y$.
The edge $E$ is emanating rightward from the vertex $( S_1(C_0)- S_1(C_1), S_1(C_0))$ of $K_1$. 
Since $ S_1(C_1)=C_1(J_1,\infty,\ldots,\infty)$, we find $C_1\leq S_1(C_1)$.
This implies $C_0-C_1\geq S_1(C_0)- S_1(C_1)$.
Thus the vertex $V_g^\prime$ of $\Gamma$ is on $E$.
For an odd number $g$ the statement is similarly shown.
\qed

\begin{figure}[htbp]
\centering
{\unitlength=.05in{
\begin{picture}(80,52)(-5,22)
\thicklines
\dottedline(55,46)(46,44)
\dottedline(55,50)(46,52)
\dottedline(55,46)(55,50)
\dottedline(46,44)(37,41)
\dottedline(46,52)(37,55)
\dottedline(46,44)(46,52)
\dottedline(37,41)(37,55)
\dashline{1}(37,41)(27,36)
\dashline{1}(37,55)(27,60)
\dottedline(27,36)(27,60)
\dottedline(27,36)(21,32)
\dottedline(27,60)(21,64)
\dottedline(21,32)(21,64)
\dottedline(21,32)(13,24)
\dottedline(21,64)(13,72)
\dottedline(13,24)(13,72)
\put(11,20){\line(1,2){5}}
\put(16,30){\line(0,1){43}}
\put(16,30){\line(1,1){3}}
\put(19,33){\line(0,1){40}}
\put(19,33){\line(3,2){6}}
\put(25,37){\line(0,1){36}}
\dashline{1}(25,37)(39,43)
\put(39,43){\line(0,1){30}}
\put(39,43){\line(4,1){12}}
\put(51,46){\line(0,1){27}}
\put(51,46){\line(1,0){8}}
\put(55,43){\makebox(0,0){$V_g^\prime$}}
\put(10,25){\makebox(0,0){$V_0^\prime$}}
\end{picture}
}}
\caption{
The tropical curve drown by solid lines is $K_1$ for an odd number $g$. 
The curve drown by dotted lines is the tropical hyperelliptic curve $\Gamma$.
}\label{fig:K1}
\end{figure}
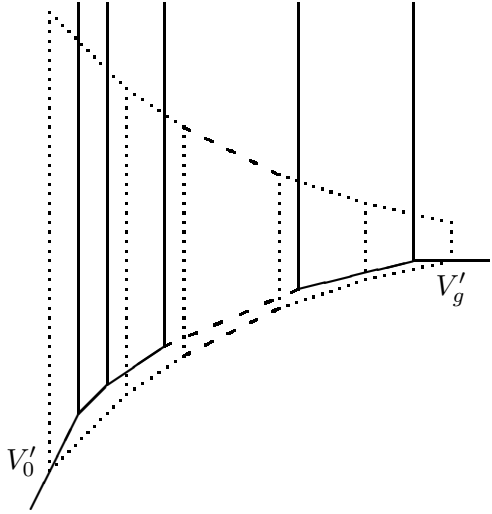

Figure \ref{fig:K1} shows the tropical curves $K_1$ and $\Gamma$.
If $g$ is an even number the vertical edges of $K_1$ are given by $X=S_1(C_i)-S_1(C_{i+1})$ for $i=0,1,\ldots,g-1$.
If $g$ is an odd number the vertical edges of $K_1$ are given by $X=T_{g+1}(C_i)-T_{g+1}(C_{i+1})$ for $i=0,1,\ldots,g-1$ and $X=T_{g+1}(C_g)$.

If we assume that $g$ is an odd number then $K_1$ passes through the vertex $V_0^\prime$ because $C_{g}\leq T_{g+1}(C_g)\leq C_{g-1}-C_g$ holds.
Moreover, there exists at least one intersection point of $\Gamma$ and $K_1$ on the upper half of $\alpha_1\setminus\alpha_{1,2}$.

Let us consider the addition formula on ${\rm Sym}^g(\Gamma)$ equivalent to the time evolution of the UD-pTL (see theorem \ref{thm:tepBBSonSymg}):
\begin{align}
d_{\bar P}
=
d_P\oplus\tau
\quad
\Longleftrightarrow
\quad
-d_{\bar P}
\oplus
d_P
\oplus
\tau
=
0.
\label{eq:addtropsym}
\end{align}
Since $\phi(T)\equiv V_g^\prime-V_0^\prime$ (mod $\mathcal{D}_l(\Gamma)$) for any $g$ (see proposition \ref{prop:phiT}), the addition formula (\ref{eq:addtropsym}) is written by the divisors on ${\rm Pic}^0(\Gamma)$:
\begin{align}
D_Q+D_P+V_g^\prime-V_0^\prime-2D^\ast
\equiv
0
\quad
\mbox{(mod $\mathcal{D}_l(\Gamma)$)},
\label{eq:addtroppic}
\end{align}
where $D_Q=\mu^{-1}(-d_{\bar P})$.
Hereafter we assume that the divisor $D_P$ is in $\tilde{\mathcal{D}}$.
We have the following proposition.

\begin{proposition}\label{prop:troph1}
Assume that the points $P_1,P_2,\ldots,P_g\in\Gamma$ are on the tropical curve $K_1$.
Then the rational function $H_1|\Gamma$ on $\Gamma$ satisfies
\begin{align}
(H_1|\Gamma)
=
D_Q+D_P+V_g^\prime-V_0^\prime-2D^\ast.
\label{eq:Hpd}
\end{align}
\end{proposition}

(Proof)\quad
By lemma \ref{lem:troph2}, the curve $K_1$ passes through $V_g^\prime$. 
By the assumption of proposition, $K_1$ passes through $P_1,P_2,\ldots,P_g$ as well. 
These fact suggests that the rational function $H_1|\Gamma$ on $\Gamma$ has $g+1$ zeros at $P_1,P_2,\ldots,P_g,V_g^\prime$.
Moreover, by lemma \ref{lem:troph}, $H_1|\Gamma$ is in the maximal tropical module $\mathcal{M}_1$ in $L(V_0^\prime+2D^\ast)$. 
Therefore, $H_1|\Gamma$ satisfies (\ref{eq:Hpd}) and is uniquely determined up to an additive constant.
\qed

Thus the points $P_1,P_2,\ldots,P_g$, $Q_1,Q_2,\ldots,Q_g$, $V_g^\prime$ satisfying (\ref{eq:addtroppic}) are on both $\Gamma$ and $K_1$.
Therefore, the addition formula (\ref{eq:addtropsym}) can be realized by using the intersection of $\Gamma$ and $K_1$.
Note that we have $D_Q\in\tilde{\mathcal{D}}$ because $D_P,D_T\in\tilde{\mathcal{D}}$ and $\tilde{\mathcal{D}}\simeq{\rm Pic}^0(\Gamma)$.
In order to realize the time evolution of the UD-pTL by using tropical curve intersections, we have only to give the inverse $d_{\bar P}$ of $-d_{\bar P}=\mu(D_Q)$ in terms of a tropical curve intersection.

\subsection{Inverse elements}
Consider the addition formula on ${\rm Sym}^g(\Gamma)$
\begin{align*}
-d_{\bar P}
\oplus
d_{\bar P}
=
o.
\end{align*}
This can be written by the divisors on ${\rm Pic}^0(\Gamma)$:
\begin{align*}
D_Q
+
D_{\bar P}
-
2D^\ast
\equiv
0
\quad
\mbox{(mod $\mathcal{D}_l(\Gamma)$).}
\end{align*}
Therefore, the tropical curve whose intersection with $\Gamma$ gives the inverse $d_{\bar P}$ of $-d_{\bar P}=\mu(D_Q)$ is given by a rational function in $L(2D^\ast)$.
In the following, we show that the ultradiscretization of the rational function $h_2$ gives such a tropical curve.

Assume that $g$ is an odd number.
Consider the curve $\kappa_2=\left(h_2(u,v)=0\right)$ introduced in \S\ref{sec:invelntp}.
Applying $\sigma$ to the points on $\kappa_2$, we obtain an affine curve 
\begin{align*}
{\tilde\kappa_2}
:=
\left(
\tilde h_2(x,y)
=
0
\right)
=
\left\{
(x,y)
=
\sigma(u,v)\ \left|\
h_2(u,v)=0
\right.
\right\},
\end{align*}
where 
\begin{align*}
\tilde h_2(x,y)
&=
\sum_{i=0}^{g-1}\left(c_{i}+s_1(c_i)+2I_1\acute c_{i+1}\right)x^i
+
2c_gx^g
+
2x^{g+1}
+
2y.
\end{align*}
Remember that $\acute c_i$ is obtained from $c_i$ by setting $I_1=V_1=0$ for $i=1,2,\ldots,g$.
Since the polynomial $\tilde h_2$ is subtraction-free, by applying the procedure of ultradiscretization, we obtain the tropical polynomial
\begin{align*}
H_2(X,Y)
=
\left\lfloor
\acute C_{0},
\acute C_{1}+X,
\cdots,
\acute C_{g-1}+(g-1)X,
C_g+gX,
(g+1)X,
Y
\right\rfloor,
\end{align*}
where $\acute C_i=\lim_{\epsilon\to0}-\epsilon\log(c_{i}+ s_1(c_i)+2I_1\acute c_{i+1})$ for $i=0,1,\ldots,g-1$.
In order to consider non-degenerate curves, we assume $\acute C_i+\acute C_{i+2}>2\acute C_{i+1}$ for $i=0,1,\ldots,g-3$. 

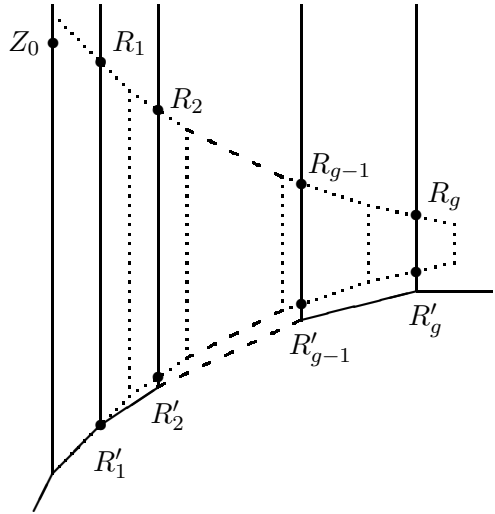
\begin{figure}[htbp]
\centering
{\unitlength=.05in{\def\arraystretch{1.0}
\begin{picture}(80,52)(-5,22)
\thicklines
\dottedline(55,46)(46,44)
\dottedline(55,50)(46,52)
\dottedline(55,46)(55,50)
\dottedline(46,44)(37,41)
\dottedline(46,52)(37,55)
\dottedline(46,44)(46,52)
\dottedline(37,41)(37,55)
\dashline{1}(37,41)(27,36)
\dashline{1}(37,55)(27,60)
\dottedline(27,36)(27,60)
\dottedline(27,36)(21,32)
\dottedline(27,60)(21,64)
\dottedline(21,32)(21,64)
\dottedline(21,32)(13,24)
\dottedline(21,64)(13,72)
\dottedline(13,24)(13,72)
\put(11,20){\line(1,2){2}}
\put(13,24){\line(0,1){49}}
\put(13,24){\line(1,1){5}}
\put(18,29){\line(0,1){44}}
\put(18,29){\line(3,2){6}}
\put(24,33){\line(0,1){40}}
\dashline{1}(24,33)(39,40)
\put(39,40){\line(0,1){33}}
\put(39,40){\line(4,1){12}}
\put(51,43){\line(0,1){30}}
\put(51,43){\line(1,0){8}}
\put(18,29){\circle*{1}}
\put(13,69){\circle*{1}}
\put(18,67){\circle*{1}}
\put(24,62){\circle*{1}}
\put(24,34){\circle*{1}}
\put(39,54.3){\circle*{1}}
\put(39,41.7){\circle*{1}}
\put(51,51){\circle*{1}}
\put(51,45){\circle*{1}}
\put(19,25){\makebox(0,0){$R_1^\prime$}}
\put(10,69){\makebox(0,0){$Z_0$}}
\put(21,69){\makebox(0,0){$R_1$}}
\put(27,63){\makebox(0,0){$R_2$}}
\put(25,30){\makebox(0,0){$R_2^\prime$}}
\put(43,56){\makebox(0,0){$R_{g-1}$}}
\put(41,37){\makebox(0,0){$R_{g-1}^\prime$}}
\put(54,53){\makebox(0,0){$R_{g}$}}
\put(52,40){\makebox(0,0){$R_{g}^\prime$}}
\end{picture}
}}
\caption{
The tropical curve $K_2$ defined by $H_2$ is drown by solid lines. 
The intersection points of $K_2$ and the tropical hyperelliptic curve $\Gamma$, drown by dotted lines, are $R_i$ and $R_i^\prime$ for $i=1,2,\ldots,g$.
The point $Z_0$ satisfies the condition (\ref{eq:conz0}).
}
\label{fig:K2}
\end{figure}

We define the tropical curve $K_2$ to be the set of points at which $H_2$ is not differentiable:
\begin{align*}
K_2
:=
\left\{
P\in\R^2\ |\ 
\mbox{$H_2$ is not differentiable at $P$}
\right\}.
\end{align*}
Figure \ref{fig:K2} shows the tropical curve $K_2$.

Note that the terms $(g+1)X$ and $Y$ are never dominant in $H_2$ on $\Gamma$.
Therefore, the restriction of $H_2$ on $\Gamma$, which is denoted by $H_2|\Gamma$, is given by the formula
\begin{align*}
H_2|\Gamma(X,Y)
=
\left\lfloor
\acute C_{0},
\acute C_{1}+X,
\cdots,
\acute C_{g-1}+(g-1)X,
C_g+gX
\right\rfloor.
\end{align*}

On can show that $\acute C_{g-1}\leq C_{g-1}$ holds for any odd number $g$.
Thus we can let $R_1,R_1^\prime,\ldots,R_g,R_g^\prime$ be the intersection points of $\Gamma$ and $K_2$, where we assume that $R_i$ is on the upper half of $\Gamma$ (see figure \ref{fig:K2}).

The rational function $H_2|\Gamma$ on $\Gamma$ has poles of oder $g$ at $V_0$ and $V_0^\prime$.
This implies
\begin{align}
\left(
H_2|\Gamma
\right)
=
D_R+D_{R^\prime}-g(V_0+V_0^\prime)
=
D_R+D_{R^\prime}+V_0-V_0^\prime-2D^\ast,
\label{eq:rrpeq}
\end{align}
where $2D^\ast=(g+1)V_0+(g-1)V_0^\prime$. 
There uniquely exists a point $Z_0$ on $\alpha_1\setminus\alpha_{1,2}$ such that $V_0+R_1^\prime-V_0^\prime-Z_0$ is the principal divisor of a rational function on $\Gamma$ (see figure \ref{fig:K2}).
The point $Z_0$ satisfies the condition
\begin{align}
Q(\overline{V_0^\prime R_1^\prime},\overline{V_0^\prime R_1^\prime})
=
Q(\overline{V_0Z_0},\overline{V_0Z_0}).
\label{eq:conz0}
\end{align}
We define $\tilde H_2$ to be the rational function on $\Gamma$ whose principal divisor is $D_R+Z_0+R_2^\prime+\cdots+R_g^\prime-2D^\ast$.
Then we find $(\tilde H_2)\equiv(H_2|\Gamma)$ (mod $\mathcal{D}_l(\Gamma)$) and $\tilde H_2\in L(2D^\ast)$.

\begin{proposition}\label{prop:invelud}
Let $D_P$ and $D_Q$ be elements of $\tilde{\mathcal{D}}$ satisfying (\ref{eq:Hpd}).
Also let $D_R$ and $D_{R^\prime}$ be elements of $\tilde{\mathcal{D}}$ satisfying (\ref{eq:rrpeq}).
Then, for any odd number $g$, we have 
\begin{align*}
D_Q
+
Z_1
+
Q_2^\prime
+
\cdots
+
Q_g^\prime
-
2D^\ast
\equiv
0
\quad
(\mbox{mod $\mathcal{D}_l(\Gamma)$}),
\end{align*}
where $Z_1$ is the unique point on $\alpha_1\setminus\alpha_{1,2}$ such that $Z_1+Q_1-Z_0-R_1$ is the principal divisor of a rational function on $\Gamma$.
Moreover, we have $Z_1+Q_2^\prime+\cdots+Q_g^\prime\in\tilde{\mathcal{D}}$.
\end{proposition}

(Proof)\quad
Since there exists at least one intersection point of $\Gamma$ and $K_1$ on the upper half of $\alpha_1\setminus\alpha_{1,2}$, we can assume that $Q_1$ is the one.
Then there exist rational functions $G_1$ and $G_2$ on $\Gamma$ such that
\begin{align*}
\left(G_1\right)
&=
Z_1+Q_1-Z_0-R_1,\\
\left(G_2\right)
&=
Q_2+\cdots+Q_g+Q_2^\prime+\cdots+Q_g^\prime-R_2-\cdots-R_g-R_2^\prime-\cdots-R_g^\prime.
\end{align*}
Let $H$ be the rational function $G_1+G_2$ on $\Gamma$.
Then  we have
\begin{align*}
(H)
&=
Z_1
+
D_Q
+
Q_2^\prime
+
\cdots
+
Q_g^\prime
-
Z_0
-
D_R
-
R_2^\prime-\cdots-R_g^\prime.
\end{align*}
Note that we also have
\begin{align*}
\left(\tilde H_2\right)
=
D_R+Z_0+R_2^\prime+\cdots+R_g^\prime-2D^\ast.
\end{align*}
We then find
\begin{align*}
\left(H+\tilde H_2\right)
=
Z_1
+
D_Q
+
Q_2^\prime
+
\cdots
+
Q_g^\prime
-
2D^\ast
\equiv
0
\quad
(\mbox{mod $\mathcal{D}_l(\Gamma)$})
\end{align*}
and $H+\tilde H_2\in L(2D^\ast)$.
It is clear that $Z_1+Q_2^\prime+\cdots+Q_g^\prime\in\tilde{\mathcal{D}}$.
\qed

If $g$ is an even number we have (see \cite{Nobe12})
\begin{align*}
D_{\bar P}
=
D_{Q^\prime}.
\end{align*}
If $g$ is an odd number proposition \ref{prop:invelud} implies
\begin{align*}
D_{\bar P}
=
Z_1+Q_2^\prime+\cdots+Q_g^\prime.
\end{align*}
Thus we obtain the following theorem.

\begin{theorem}
Let $D_P$ and $D_Q$ be elements of $\tilde{\mathcal{D}}$ satisfying (\ref{eq:Hpd}).
Also let $T$ be the element of $\tilde{\mathcal{D}}$ given by (\ref{eq:tepBBS}).
Put $d_{P}=\mu(D_P)\in{\rm Sym}^g(\Gamma)$ and $\tau=\mu(T)\in{\rm Sym}^g(\Gamma)$.
Then the element $d_{\bar P}\in{\rm Sym}^g(\Gamma)$ defined by the addition 
\begin{align*}
d_{\bar P}
=
d_P\oplus\tau
\end{align*}
is explicitly given by the formula
\begin{align*}
d_{\bar P}
=
\begin{cases}
\left\{
Q_1^\prime,Q_2^\prime,\ldots,Q_g^\prime
\right\}
&\mbox{for even $g$},\\
\left\{
Z_1,Q_2^\prime,\ldots,Q_g^\prime
\right\}
&\mbox{for odd $g$.}\\
\end{cases}
\end{align*}
\qed
\end{theorem}

Thus the time evolution of the UD-pTL is realized by using the intersection of the tropical curves $\Gamma$, $K_1$, and $K_2$.

\begin{example}
Put $g=3$.
The spectral curve $\Gamma$ is given by 
\begin{align*}
&F(X,Y)
=
\left\lfloor
2X,
Y+\left\lfloor
4X,C_3+3X,C_2+2X,C_1+X,C_0
\right\rfloor,
C_{-1}
\right\rfloor,\\
&C_3
=
\left\lfloor
J_i,W_i
\right\rfloor_{1\leq i\leq 4},\quad
C_2
=
\left\lfloor
\underset{1\leq i<j\leq4}
{\left\lfloor
J_i+J_j,W_i+W_j
\right\rfloor},
\underset{1\leq i\leq4}
{\left\lfloor
J_i+W_{i+1},J_i+W_{i+2}
\right\rfloor}
\right\rfloor,\\
&C_1
=
\left\lfloor
\underset{1\leq i<j<k\leq4}
{\left\lfloor
J_i+J_j+J_k,W_i+W_j+W_k
\right\rfloor},
\underset{1\leq i\leq4}
{\left\lfloor
J_i+W_{i+2}+
\left\lfloor
J_{i+1},W_{i+1}
\right\rfloor
\right\rfloor}
\right\rfloor,\\
&C_0
=\sum_{i=1}^4J_i,\quad
C_{-1}
=\sum_{i=1}^4\left(J_i+W_i\right).
\end{align*}

The tropical curves $K_1$ and $K_2$ are respectively given by 
\begin{align*}
H_1(X,Y)
=
\left\lfloor
Y,4X,
T_4(C_3)+3X,T_4(C_2)+2X,T_4(C_1)+X,T_4(C_0)
\right\rfloor,
\\
H_2(X,Y)
=
\left\lfloor
Y,4X,
C_3+3X,\acute C_2+2X,\acute C_1+X,\acute C_0
\right\rfloor.
\end{align*}

Figure \ref{fig:hecg3} shows an example of the intersection of $\Gamma$, $K_1$, and $K_2$.

\begin{figure}[htbp]
\centering
{\unitlength=.08in{
\begin{picture}(10,40)(9,-2)
\thicklines
\put(0,0){\line(0,1){36.5}}
\put(0,0){\line(4,3){14}}
\put(0,36.5){\line(4,-3){14}}
\put(14,10.5){\line(0,1){15.5}}
\put(14,10.5){\line(2,1){10}}
\put(14,26){\line(2,-1){10}}
\put(24,15.5){\line(0,1){5.5}}
\put(24,15.5){\line(4,1){6}}
\put(24,21){\line(4,-1){6}}
\put(30,17){\line(0,1){2.5}}
\dashline{1}(8,8)(-2,-2)
\dashline{1}(8,8)(8,37)
\dashline{1}(8,8)(12,11)
\dashline{1}(12,11)(12,37)
\dashline{1}(12,11)(20,15)
\dashline{1}(20,15)(20,37)
\dashline{1}(20,15)(28,17)
\dashline{1}(28,17)(28,37)
\dashline{1}(28,17)(32,17)
\dottedline{.3}(12,9)(-2,-1.5)
\dottedline{.3}(12,9)(12,37)
\dottedline{.3}(12,9)(20,13)
\dottedline{.3}(20,13)(20,37)
\dottedline{.3}(20,13)(28,15)
\dottedline{.3}(28,15)(28,37)
\dottedline{.3}(28,15)(32,15)
\put(0,33){\circle*{.5}}
\put(0,34.5){\circle*{.5}}
\put(8,30.5){\circle*{.5}}
\put(20,13.5){\circle*{.5}}
\put(28,16.5){\circle*{.5}}
\put(14,12){\circle*{.5}}
\put(14,26){\circle*{.5}}
\put(24,16){\circle*{.5}}
\put(30,17){\circle*{.5}}
\put(-2,35){\makebox(0,0){$Z_1$}}
\put(-2,33){\makebox(0,0){$B$}}
\put(10,31){\makebox(0,0){$P_1$}}
\put(15.5,15){\makebox(0,0){$P_2$}}
\put(21,11.5){\makebox(0,0){$Q_2^\prime$}}
\put(25.5,18){\makebox(0,0){$P_3$}}
\put(28.5,13.5){\makebox(0,0){$Q_3^\prime$}}
\put(31.5,18.5){\makebox(0,0){$V_3^\prime$}}
\put(15,27.5){\makebox(0,0){$V_1$}}
\end{picture}
}}
\caption{
The solid curve is $\Gamma$, the dashed one is $K_1$, and the dotted one is $K_2$.
The set $\{P_1,P_2,P_3\}$ of points is mapped into $\{Z_1,Q_2^\prime,Q_3^\prime\}$ by the addition by $\{B,V_1,V_3^\prime\}$ on ${\rm Sym}^3(\Gamma)$.
This map is equivalent to the time evolution of the UD-pTL.
Here we set $J_1=3,J_2=5,J_3=7,J_4=2,W_1=3.5,W_2=6,W_3=10,W_4=0$.
}
\label{fig:hecg3}
\end{figure}
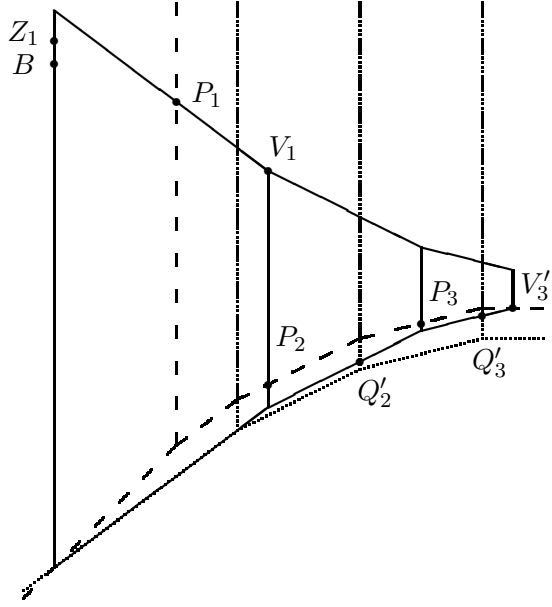
\end{example}

\subsection{Discrete motion of tropical curves}
We can interpret the time evolution of the UD-pTL as discrete motion of tropical curves.

Let $\mathcal{C}_1$ be the moduli space of the tropical curve $K_1$
\begin{align*}
\mathcal{C}_1
=
\begin{cases}
\left\{\left(S_1(C_0),S_1(C_1),\ldots,S_1(C_g)\right)\right\}&\mbox{for even $g$,}\\
\left\{\left(T_{g+1}(C_0),T_{g+1}(C_1),\ldots,T_{g+1}(C_g)\right)\right\}&\mbox{for odd $g$.}\\
\end{cases}
\end{align*}
Define a map $\tilde\psi: \mathcal{T}_C\to\mathcal{C}_1$ to be
\begin{align*}
\tilde\psi: (J_1,\ldots,J_{g+1},W_1,\ldots,W_{g+1})
\mapsto
\begin{cases}
\left(S_1(C_0),\ldots,S_1(C_g)\right)&\mbox{for even $g$,}\\
\left(T_{g+1}(C_0),\ldots,T_{g+1}(C_g)\right)&\mbox{for odd $g$.}\\
\end{cases}
\end{align*}
Also define a map $\upsilon$ on $\mathcal{C}_1$ to be
\begin{align*}
\upsilon: 
\begin{cases}
\left(S_1(C_0),\ldots,S_1(C_g)\right)
\mapsto
\left(S_1(\bar C_0),\ldots,S_1(\bar C_g)\right)&\mbox{for even $g$,}\\
 \left(T_{g+1}(C_0),\ldots,T_{g+1}(C_g)\right)
\mapsto
\left(T_{g+1}(\bar C_0),\ldots,T_{g+1}(\bar C_g)\right)&\mbox{for odd $g$}\\
\end{cases}
\end{align*}
so that the following diagram is commutative
\begin{align*}
\begin{CD}
\mathcal{T}_C @> \tilde\psi >> \mathcal{C}_1\\
@V (\ref{eq:pbbs}) VV @VV \upsilon V\\
\mathcal{T}_{C} @>> \tilde\psi > \mathcal{C}_1.\\
\end{CD}
\end{align*}

Then $\upsilon$ induces the discrete motion of tropical curves
\begin{align*}
K_1^0\to K_1^1\to K_1^2\to\cdots,
\end{align*}
where $K_1^t$ is given by the tropical polynomial
\begin{align*}
&H_1^t(X,Y)
=
\begin{cases}
\left\lfloor
\underset{0\leq i\leq g}{\left\lfloor S_1(C_i^t)+iX\right\rfloor},Y
\right\rfloor
&\mbox{for even $g$,}\\
\left\lfloor
\underset{0\leq i\leq g}{\left\lfloor T_{g+1}(C_i^t)+iX\right\rfloor},(g+1)X,Y
\right\rfloor
&\mbox{for odd $g$.}
\end{cases}
\end{align*}

\begin{figure}[t]
\centering
{
\includegraphics[scale=.4]{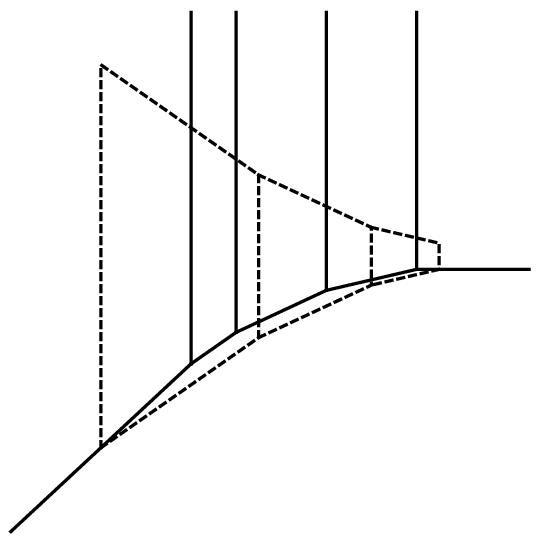}
\includegraphics[scale=.4]{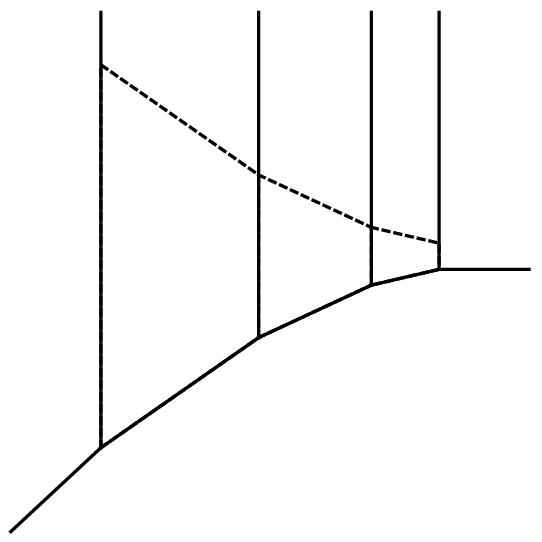}
\includegraphics[scale=.4]{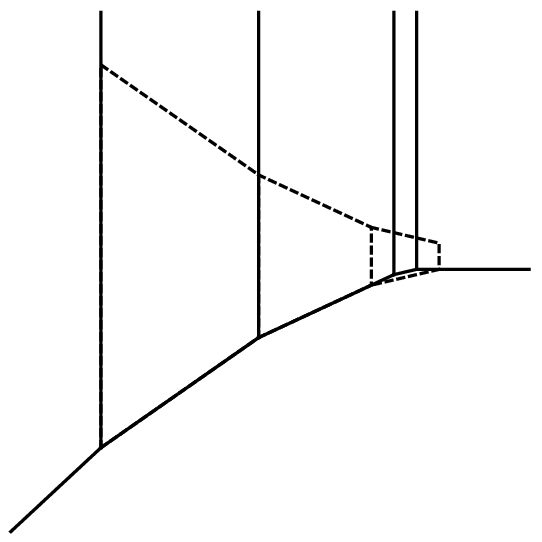}
\includegraphics[scale=.4]{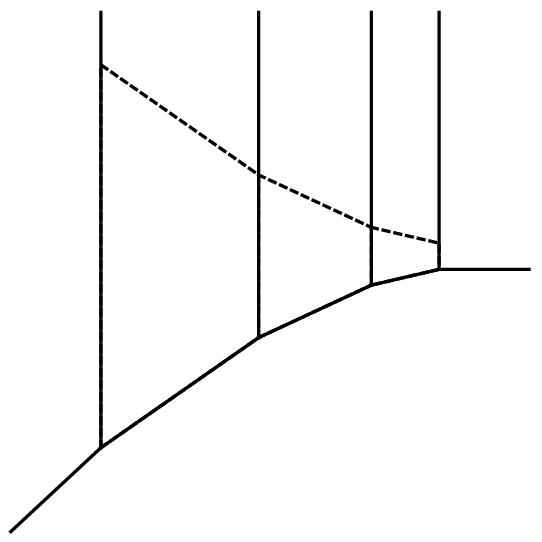}
\\
\includegraphics[scale=.4]{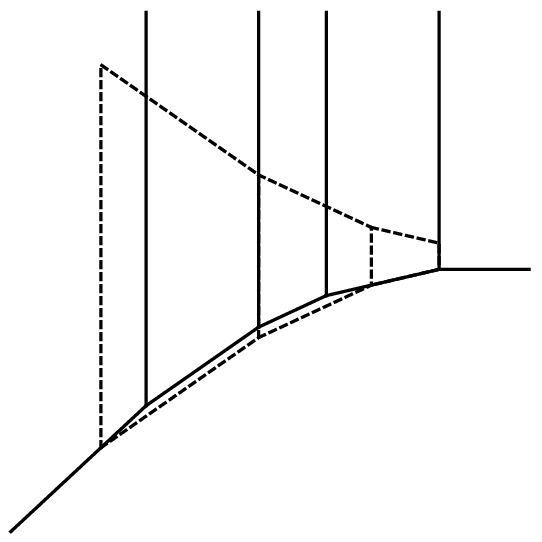}
\includegraphics[scale=.4]{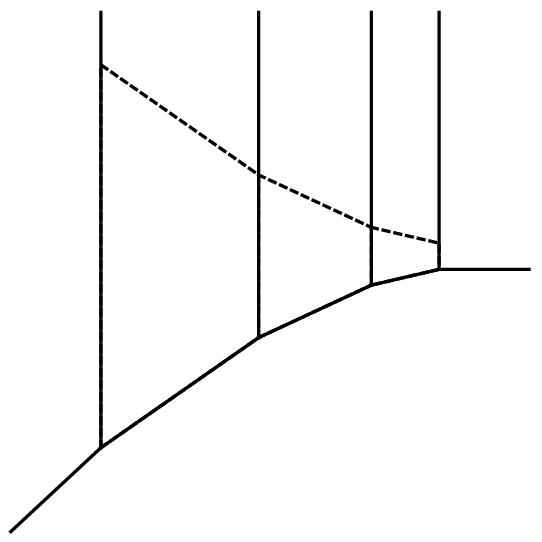}
\includegraphics[scale=.4]{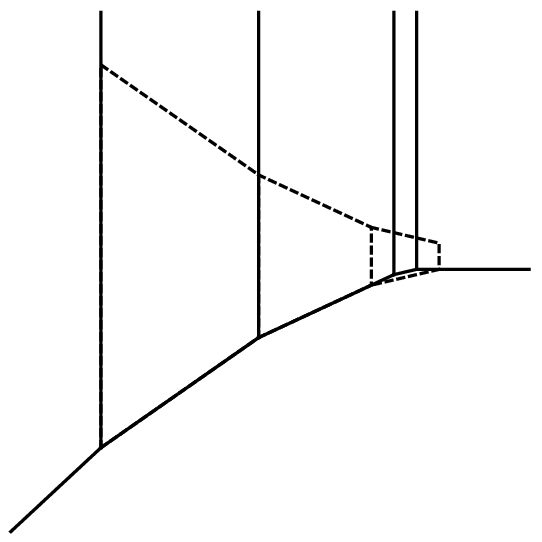}
\includegraphics[scale=.4]{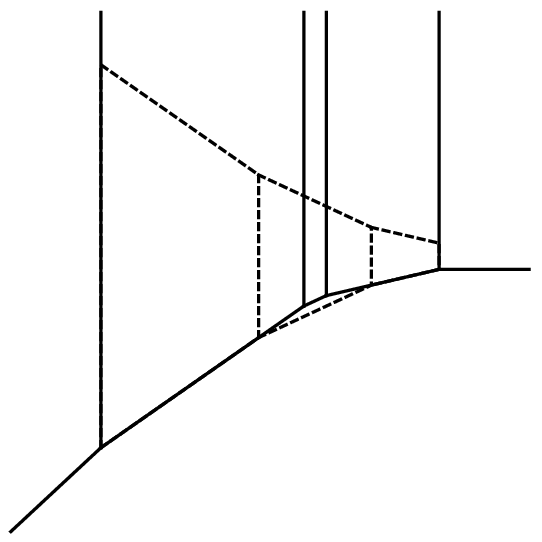}
}
\caption{The discrete motion of the tropical curves $K_1^t$ ($t=0,1,\ldots,12$) induced by $\upsilon$. 
The solid curves are $K_1^t$ and the dashed one is $\Gamma$.
The figures are sorted in time increasing order from left to right and top to bottom.
Note that $K_1^1=K_1^2=\cdots=K_1^6$ and the second figure shows them.
(This fact does not suggest that the intersection points of $\Gamma$ and $K_1^t$ ($t=1,2,\ldots,6$) are fixed.)
Here we set $J_1^0=3,J_2^0=5,J_3^0=7,J_4^0=2,W_1^0=3.5,W_2^0=6,W_3^0=10,W_4^0=0$.
}
\label{fig:TCMG3}
\end{figure}

Figure \ref{fig:TCMG3} shows an example of the discrete motion of $K_1^t$.

\section{Concluding remarks}
\label{sec:CONCL}
We establish a geometric realization of the pdTL via the curve intersections of its spectral curve $\gamma$ and two affine curves $\kappa_1$ and $\kappa_2$.
Namely, the linear flow on the Jacobian $J(\gamma)$ equivalent to the time evolution of the pdTL is translated into the curve intersections of $\gamma$, $\kappa_1$, and $\kappa_2$.
The rational functions $f$, $h_1$, and $h_2$ which respectively define $\gamma$, $\kappa_1$, and $\kappa_2$ are explicitly given by using the conserved quantities $c_{-1},c_0,\ldots,c_g$ of the pdTL.
In addition, these rational functions can simultaneously be mapped into subtraction-free ones by the rational transformation $\sigma$ on $\C^2$. 
Therefore, we can naturally apply the procedure of ultradiscretization to them and obtain the tropical hyperelliptic curve $\Gamma$ and two tropical curves $K_1$ and $K_2$ intersecting each other.
The tropical hyperelliptic curve $\Gamma$ thus obtained is nothing but the spectral curve of the UD-pTL. 
Hence, the coefficients $C_{-1},C_0,\ldots,C_g$ of its defining tropical polynomial $F$ are the conserved quantities of the UD-pTL, and the time evolution of the UD-pTL is linearized on its tropical Jacobian $J(\Gamma)$.
Moreover, two tropical curves $K_1$ and $K_2$ thus obtained are explicitly given by using the conserved quantities of the UD-pTL. 
We then show that the tropical curve intersections of $\Gamma$, $K_1$, and $K_2$ give the linear flow on $J(\Gamma)$ equivalent to the time evolution of the UD-pTL.
Thus we also establish a tropical geometric realization of the UD-pTL via the tropical curve intersections of $\Gamma$, $K_1$, and $K_2$.

It should be noted that the time evolutions of the pdTL and UD-pTL lead to discrete motions of (tropical) curves $\kappa_1$ and $K_1$ which appear in their (tropical) geometric realizations, respectively.
Because the coefficients of the defining polynomial $h_1$ (resp. $H_1$) depend on time, while those of the defining polynomial $f$ (resp. $F$) of the spectral curve $\gamma$ (resp. $\Gamma$) do not.
The time dependence of the moving curves $\kappa_1$ and $K_1$ is derived from a simple action on the phase space $\mathcal{U}$ of the pdTL which changes the sign of $I_1$ or $V_{g+1}$.
Thus the moduli spaces of $\kappa_1$ and $K_1$ are easily obtained from that of $\gamma$ and $\Gamma$ by applying such actions, respectively.
We do not know the reason why we can obtain these moduli spaces such a way.
To investigate the time evolutions of the pdTL and UD-pTL as actions on the moduli spaces of the (tropical) hyperelliptic curves is a further problem.

\section*{Acknowledgments}
This work was partially supported by JSPS KAKENHI Grant Number 22740100.

\appendix
\section{Proof of theorem \ref{thm:ker}}
\label{app:proof1}
Assume that $g$ is an even number.
If $D_P=P_1+P_2+\cdots+P_g\in\ker \Phi$ there exists a rational function $h\in L\left(D^\ast\right)$ on $H$ such that $(h)=D_P-D^\ast$.
By the Riemann-Roch theorem, we have
\begin{align*}
\dim L(0)
=
-g+1
+
\dim L\left(W\right).
\end{align*}
Noting $\dim L(0)=1$, we obtain $\dim L\left(W\right)=g$, where $W$ is a canonical divisor on $H$.
Now choose $W=\left(dx/y\right)=(g-1)\left(P_\infty+P_\infty^\prime\right)$.
Again, by the Riemann-Roch theorem, we have
\begin{align*}
\dim L(P_\infty+P_\infty^\prime)
=
-g+3
+
\dim L\left((g-2)\left(P_\infty+P_\infty^\prime\right)\right).
\end{align*}
Let $x$ be the coordinate function:
\begin{align*}
x(P)=p
\quad
\mbox{for $P=(p,q)\in H$}.
\end{align*}
We then have $(x)=P_0+P_0^\prime-P_\infty-P_\infty^\prime$, where $P_0$ is the point whose $x$-component is 0. 
This implies $\langle1,x\rangle\subset L(P_\infty+P_\infty^\prime)$.
It immediately follows $\dim L(P_\infty+P_\infty^\prime)\geq2$.
Therefore we obtain
\begin{align*}
\dim L\left((g-2)\left(P_\infty+P_\infty^\prime\right)\right)
\geq
g-1.
\end{align*}
Moreover, since $L\left((g-2)\left(P_\infty+P_\infty^\prime\right)\right)\subsetneq L\left(W\right)$ and $\dim L\left(W\right)=g$, we find
\begin{align*}
\dim L\left((g-2)\left(P_\infty+P_\infty^\prime\right)\right)=g-1.
\end{align*}
We inductively obtain
\begin{align*}
\dim L\left(i\left(P_\infty+P_\infty^\prime\right)\right)=i+1
\qquad
(i=1,2,\ldots,g-1).
\end{align*}
In particular, for $i=g/2$, we have
\begin{align*}
\dim L\left(D^\ast\right)
=
\dim L\left(\frac{g}{2}\left(P_\infty+P_\infty^\prime\right)\right)=\frac{g}{2}+1.
\end{align*}

On the other hand, since $(x^i)=i\left(P_0+P_0^\prime-P_\infty-P_\infty^\prime\right)$, we have
\begin{align*}
\left\langle
1,x,x^2,\cdots,x^\frac{g}{2}
\right\rangle
\subset
L\left(D^\ast\right)
\end{align*}
and 
\begin{align*}
\dim \left\langle1,x,x^2,\cdots,x^\frac{g}{2}\right\rangle=\frac{g}{2}+1.
\end{align*}
Therefore, we obtain
\begin{align*}
\left\langle
1,x,x^2,\cdots,x^\frac{g}{2}
\right\rangle
=
L\left(D^\ast\right).
\end{align*}

Thus a rational function $h\in L(D^\ast)$ can be expressed as follows
\begin{align*}
h=a_0+a_1x+\cdots+a_{\frac{g}{2}}x^{\frac{g}{2}}
\qquad
(a_i\in\C).
\end{align*}
If $a_{{g}/{2}}\neq0$ the equation $h=0$ has exactly $g/2$ solutions $c_1,c_2,\cdots,c_{\frac{g}{2}}$, counting multiplicities.
Thus the principal divisor of $h$ is 
\begin{align*}
(h)
=
(x-c_1)
+
(x-c_2)
+
\cdots
+
(x-c_{\frac{g}{2}}).
\end{align*}
Noting
\begin{align*}
(x-c_i)
=
P_{c_i}
+
P_{c_i}^\prime
-
P_\infty
-
P_\infty^\prime,
\end{align*}
where $P_{c_i}$ is the point whose $x$-component is $c_i$, we have
\begin{align*}
(h)
=
\sum_{i=1}^{\frac{g}{2}}
\left(
P_{c_i}
+
P_{c_i}^\prime
\right)
-
D^\ast.
\end{align*}
Therefore, we obtain $D_P=\sum_{i=1}^{\frac{g}{2}}\left(P_{c_i}+P_{c_i}^\prime\right)$ as desired.

If $a_{{g}/{2}}=0$ and $a_{{g}/{2}-1}\neq0$ the equation $h=0$ has exactly $g/2-1$ solutions $c_1,c_2,\cdots,c_{{g}/{2}-1}$, counting multiplicities.
Hence we have
\begin{align*}
(h)
=
\sum_{i=1}^{\frac{g}{2}-1}
\left(
P_{c_i}
+
P_{c_i}^\prime
\right)
+
P_\infty
+
P_\infty^\prime
-
D^\ast.
\end{align*}
This is equivalent to $D_P=\sum_{i=1}^{{g}/{2}-1}\left(P_{c_i}+P_{c_i}^\prime\right)+P_\infty+P_\infty^\prime$.
Thus we inductively obtain the desired result.

For an odd number $g$, the statement is similarly shown.
\qed

\section{Proof of theorem \ref{thm:ker2}}
\label{app:proof2}
If $\Phi(D_P)=\Phi(D_Q)$ we have
\begin{align}
D_P-D^\ast\equiv D_Q-D^\ast
\quad
\Longleftrightarrow
\quad
D_P-D_Q\equiv 0,
\label{eq:dpdq}
\end{align}
where the equivalence is considered modulo $\mathcal{D}_l(H)$.
Let the $x$-component of $Q_i$ be $q_i$ for $i=1,2,\ldots,g$.
We then have $(x-q_i)=Q_i+Q_i^\prime-P_\infty-P_\infty^\prime$.
Hence we obtain
\begin{align}
D_Q+D_{Q^\prime}-g\left(P_\infty+P_\infty^\prime\right)\equiv 0
\quad
\mbox{(mod $\mathcal{D}_l(H)$).}
\label{eq:dqdqp}
\end{align}
By (\ref{eq:dpdq}) and (\ref{eq:dqdqp}), we have
\begin{align}
D_P+D_{Q^\prime}-g\left(P_\infty+P_\infty^\prime\right)\equiv 0
\quad
\mbox{(mod $\mathcal{D}_l(H)$).}
\label{eq:dpdqpgpp}
\end{align}
Therefore, there exists a rational function $h$ on $H$ such that
\begin{align}
(h)
=
D_P+D_{Q^\prime}-g\left(P_\infty+P_\infty^\prime\right).
\label{eq:dpdqp}
\end{align}
Since $h\in L(g\left(P_\infty+P_\infty^\prime\right))$ and $L(g\left(P_\infty+P_\infty^\prime\right))=\left\langle 1,x,x^2,\cdots,x^g\right\rangle$ (see the proof of theorem \ref{thm:ker}), the rational function $h$ can be expressed as $h=\sum_{i=1}^ga_i x^i$ ($a_i\in\C$).

Let $j$ be the number such that
\begin{align*}
\mbox{$a_{g}=a_{g-1}=\cdots=a_{g-j+1}=0$, $a_{g-j}\neq0$, and $0\leq j\leq g$.}
\end{align*}
Then the  rational function $h$ is factorized as follows
\begin{align*}
h=\prod_{i=1}^{g-j}(x-c_i),
\end{align*}
where $c_i$ ($i=1,2,\ldots,g-j$) is the solutions of $h=0$.
The principal divisor of $h$ is
\begin{align}
(h)
=
\sum_{i=1}^{g-j}\left(P_{c_i}+P_{c_i}^\prime\right)
-
(g-j)\left(P_\infty+P_\infty^\prime\right).
\label{eq:pc}
\end{align}
Comparing (\ref{eq:dpdqp}) and (\ref{eq:pc}), we find
\begin{align*}
D_P+D_{Q^\prime}
=
\sum_{i=1}^{g-j}\left(P_{c_i}+P_{c_i}^\prime\right)+j\left(P_\infty+P_\infty^\prime\right).
\end{align*}
The right hand side can be expressed as $D_R+D_{R^\prime}$ for $D_R=\sum_{i=1}^{g-j}P_{c_i}+jP_\infty\in\mathcal{D}_g^+$.

The converse is easily shown.
\qed

\section{A geometric realization of the pdTL for \boldmath{$g=1$}}
\label{subsec:g1}
For $g=1$, we choose the Lax matrices as follows
\begin{align*}
L
=
\left(
\begin{matrix}
I_2+V_1&1-{I_1V_1}/{y}\\
I_2V_2-y&I_1+V_2\\
\end{matrix}
\right),
\qquad
M
=
\left(
\begin{matrix}
I_2&1\\
-y&I_1\\
\end{matrix}
\right).
\end{align*}
The spectral curve $\gamma$ is given by
\begin{align*}
&f(u,v)=v^2-\left(u^{2}+c_1u+c_0\right)^2+4c_{-1},\\
&c_1
=\sum_{i=1}^2\left(I_i+V_i\right),\quad
c_0
=\prod_{i=1}^2I_i+\prod_{i=1}^2V_i,\quad
c_{-1}
=\prod_{i=1}^2I_iV_i.
\end{align*}
By solving the linear equations
\begin{align*}
\varphi_1(x,y)=I_2V_2-y=0,\quad
\varphi_2(x)=x+I_1+V_2=0,
\end{align*}
we obtain the eigenvector map
\begin{align*}
P_1
=
\phi(I_1,I_2,V_1,V_2)
=
\left(
-I_1-V_2,
I_2V_2-I_1V_1
\right).
\end{align*}

The time evolution of the pdTL is given by the addition formula on ${\rm Sym}^1(\gamma)\simeq\gamma$:
\begin{align}
\bar P_1
=
P_1
\oplus
T,
\label{eq:tetlg1}
\end{align}
where $T$ is the unique point on $\gamma$ such that $T+P_\infty^\prime-A-P_\infty$ is the principal divisor of a rational function on $\gamma$ and $A=\left(0,V_1V_2-I_1I_2\right)$.

To obtain $T$ explicitly, note first that $l\in L(A+P_\infty)$ for $(l)=T+P_\infty^\prime-A-P_\infty$ and $\dim L(A+P_\infty)=2$.
We can easily see that the rational function
\begin{align}
\frac{y+\sqrt{c_0^2-4c_{-1}}}{x}+x+c_1
\label{eq:rfg1}
\end{align}
is expanded as follows
\begin{align*}
\begin{cases}
\DIS\frac{2}{t_1}+2c_1+\left(c_0+\sqrt{c_0^2-4c_{-1}}\right)t_1+o(t_1)&\mbox{at $P_\infty$,}\\
\DIS\left(-c_0+\sqrt{c_0^2-4c_{-1}}\right)t_2+o(t_2)&\mbox{at $P_\infty^\prime$,}\\
\DIS\frac{2\sqrt{c_0^2-4c_{-1}}}{t_3}+\frac{c_0c_1}{\sqrt{c_0^2-4c_{-1}}}+c_1+o(1)&\mbox{at $A$},\\
\end{cases}
\end{align*}
where $t_1$, $t_2$, and $t_3$ are the local parameters at $P_\infty$, $P_\infty^\prime$, and $A$, respectively.
The rational function (\ref{eq:rfg1}) has poles of oder 1 at $P_\infty$ and $A$, and has a zero of order 1 at $P_\infty^\prime$.
Therefore, the rational function $l$ is nothing but (\ref{eq:rfg1}).
By solving $l=0$ and $f=0$, we obtain
\begin{align*}
T
=
\left(-c_1,-\sqrt{c_0^2-4c_{-1}}\right).
\end{align*}

The addition (\ref{eq:tetlg1}) can be written by the divisors:
\begin{align*}
\left(-\bar P_1\right)
+
P_1
+
T
-
3P_\infty
\equiv
\left(-\bar P_1\right)
+
P_1
+
A
-P_\infty^\prime
-2P_\infty
\equiv
0
\quad
\mbox{(mod $\mathcal{D}_l(\gamma)$)},
\end{align*}
where $-\bar P_1$ is the inverse of $\bar P_1$.
This implies that the rational function $h$ such that $(h)=\left(-\bar P_1\right)+P_1+A-P_\infty^\prime-2P_\infty$ is in $L(P_\infty^\prime+2P_\infty)=\left\langle1,x,x^2+y\right\rangle$.
Thus we obtain the curve $\kappa_1=\left(h_1=0\right)$ passing through $P_1$ and $A$:
\begin{align*}
h_1(u,v)=\left(I_1I_2-V_1V_2\right)+\left(I_1+I_2-V_1+V_2\right)u+u^2+v.
\end{align*}
The third intersection point of $\gamma$ and $\kappa_1$ is ${Q_1}=(-I_2-V_2,I_1 V_2-I_2 V_1)$.
We successively obtain the curve $\kappa_2=\left(h_2(u,v)=0\right)$ passing through $Q_1$:
\begin{align*}
h_2(u,v)=I_1 I_2 + V_1 V_2+2 I_2V_1+u^2+c_1u+v.
\end{align*}
The second intersection point of $\gamma$ and $\kappa_{2}$ is
\begin{align*}
(-I_1-V_1,I_1 V_2-I_2 V_1)
=
(-\bar I_1-\bar V_2,\bar I_2 \bar V_2-\bar I_1 \bar V_1)
=
{\bar P_1}.
\end{align*}
Thus the time evolution of the pdTL is realized by using the intersection of the spectral curve $\gamma$ and the affine curves $\kappa_1$, $\kappa_{2}$.


\end{document}